\documentclass[aps,prc,amsfonts,preprintnumbers,superscriptaddress,nofootinbib]{revtex4}

\usepackage{amsfonts}
\usepackage{amscd}
\usepackage{bm}
\usepackage{graphicx}
\usepackage{color}
\usepackage{empheq}
\usepackage{xfrac}
\usepackage{comment}

\usepackage{graphics}
\usepackage{amsmath}
\usepackage{amssymb}
\usepackage{psfrag}
\usepackage{epsfig}
\usepackage{epsf}
\usepackage{float}
\usepackage{slashed}
\allowdisplaybreaks

\usepackage[pdftex]{hyperref}
\hypersetup{bookmarksnumbered=true,colorlinks=true,linkcolor=magenta,citecolor=blue,breaklinks=false}
\usepackage{orcidlink}

\newcommand{\fet}[1]{\mbox{\boldmath $#1$}}

\newcommand{\beq}{\begin{equation}}
\newcommand{\eeq}{\end{equation}}
\newcommand{\beqa}{\begin{eqnarray}}
\newcommand{\eeqa}{\end{eqnarray}}
\newcommand{\nn}{\nonumber \\ }

\DeclarePairedDelimiterX\Braket[2]{\langle}{\rangle}{#1 \delimsize\vert #2}

\def\e{\kern+.5ex\lower.42ex\hbox{$\scriptstyle \iota$}\kern-1.10ex e}

\numberwithin{equation}{section}
\renewcommand\theequation{\arabic{equation}}

\usepackage{color}

\begin{document}


\title{Challenging chiral EFT with tritium beta decay}

\author{D.F.~Ram\'irez~Jim\'enez\orcidlink{0000-0002-2932-1399}}
\affiliation{M.~Smoluchowski Institute of Physics, 
Faculty of Physics, Astronomy and Applied Computer Science,
Jagiellonian University,  PL-30348 Krak\'ow, Poland}
\affiliation{Doctoral School of Exact and Natural Sciences, Jagiellonian University, PL-30348 Kraków, Poland}

\author{S.~Heihoff\orcidlink{0009-0000-3641-0640}}
\affiliation{Ruhr-Universit\"at
  Bochum, Fakult\"at f\"ur Physik und Astronomie, Institut f\"ur
  Theoretische Physik II, D-44780 Bochum, Germany}

\author{J.~Golak\orcidlink{0000-0002-5210-6910}}
\affiliation{M.~Smoluchowski Institute of Physics, 
Faculty of Physics, Astronomy and Applied Computer Science,
Jagiellonian University,  PL-30348 Krak\'ow, Poland}

\author{E.~Epelbaum\orcidlink{0000-0002-7613-0210}}
\affiliation{Ruhr-Universit\"at
  Bochum, Fakult\"at f\"ur Physik und Astronomie, Institut f\"ur
  Theoretische Physik II, D-44780 Bochum, Germany}

\author{H.~Krebs\orcidlink{0000-0002-2263-0308}}
\affiliation{Ruhr-Universit\"at
  Bochum, Fakult\"at f\"ur Physik und Astronomie, Institut f\"ur Theoretische Physik II, D-44780 Bochum, Germany}

\author{P.~Reinert\orcidlink{0000-0002-1987-8775}}
\affiliation{Ruhr-Universit\"at
  Bochum, Fakult\"at f\"ur Physik und Astronomie, Institut f\"ur Theoretische Physik II, D-44780 Bochum, Germany}

\author{R.~Skibi\'nski\orcidlink{0000-0003-0806-4634}}
\affiliation{M.~Smoluchowski Institute of Physics, 
Faculty of Physics, Astronomy and Applied Computer Science,
Jagiellonian University,  PL-30348 Krak\'ow, Poland}

\author{K.~Topolnicki\orcidlink{0000-0002-9312-1842}}
\affiliation{M.~Smoluchowski Institute of Physics, 
Faculty of Physics, Astronomy and Applied Computer Science,
Jagiellonian University,  PL-30348 Krak\'ow, Poland}

\author{H.~Wita{\l}a\orcidlink{0000-0001-5487-4035}}
\affiliation{M.~Smoluchowski Institute of Physics, 
Faculty of Physics, Astronomy and Applied Computer Science,
Jagiellonian University,  PL-30348 Krak\'ow, Poland}

\date{\today}

\begin{abstract}
  We present a detailed investigation of tritium beta decay up to third
  order (N$^2$LO) in chiral effective field theory (EFT) using the LENPIC
  interactions. Unlike existing studies, we use nucleon-deuteron
  scattering observables to fix the low-energy constant $D$ that
  governs the strength of the short-range contributions to the
  exchange axial current operator and three-nucleon forces. 
  Surprisingly, the resulting parameter-free predictions for
  the tritium Gamow-Teller reduced matrix element are found to considerably
  overestimate its  empirical value.
  This result remains robust against reasonable variations of the
  pion-nucleon coupling constants and regularization scheme.
  A closer look at the size of the parameter-free long-range
  two-body contributions to the  Gamow-Teller matrix element reveals the fine-tuned
  nature this observable in chiral EFT, which may partially
  explain the observed deviation. Our results indicate a considerable
  N$^2$LO truncation uncertainty for  tritium beta decay and point towards
  large higher-order two-body corrections. More definite conclusions await a
  complete fourth-order analysis of nucleon-deuteron  scattering
  observables and tritium half-life.
\end{abstract}

\maketitle

\section{Introduction}
\label{Sec:intro}
\def\theequation{\arabic{section}.\arabic{equation}}

Nuclear beta ($\beta$) decay is a fundamental process that challenges our
understanding of weak currents and nuclear structure. A robust understanding 
of this process is a necessary prerequisite for studying neutrinoless double-$\beta$ decay,
one of the most promising candidates for searching for physics beyond
the Standard
Model \cite{Dolinski:2019nrj}.
Historically, nuclear $\beta$ decay led to
the longstanding $g_A$-quenching puzzle related to a significant
overprediction of the empirical values of the Gamow-Teller
(GT) reduced matrix element in medium-mass and heavy nuclei by
shell-model calculations \cite{Towner:1987zz}.  Recent studies using {\it ab
  initio} methods to  solve  the 
quantum $A$-body problem indicate that 
these discrepancies can, to a large extent, be resolved by taking into
account nuclear correlations and two-body  contributions to
the axial current, with no need for an {\it ad hoc} reduction of the
nucleon axial-vector coupling constant $g_A$ \cite{Suhonen:2017krv,Gysbers:2019uyb}.
Two- (and more-) body contributions to the nuclear current operators,
often referred to as meson-exchange currents (MECs), are also expected
to play an important
role in other weak processes such as muon capture, neutrino-nucleus scattering
and low-energy weak reactions of astrophysical interest.

Exchange axial currents are receiving even more attention in the context of
chiral EFT. The spontaneously broken chiral
symmetry of QCD and gauge invariance lead to an intriguing
relationship between the two-nucleon axial
current, the three-nucleon force (3NF), and the photon-induced pion
production operator off two nucleons, since their dominant short-range contributions are
parametrized by a single low-energy constant (LEC) $D$. This opens an exciting
avenue for testing chiral EFT in strong, electromagnetic and weak reactions under
different kinematical conditions, see Refs.~\cite{Hanhart:2000gp,Gardestig:2005pp,Gardestig:2006hj,Lensky:2007zc} for selected applications.
Nuclear axial current operators have been extensively studied in the
chiral EFT framework, beginning with the pioneering work of Park {\it et
al.} \cite{Park:1993jf}. As in the case of the vector current \cite{Friar:1977xh,Krebs:2019aka},
single-nucleon axial current operator can be conveniently
expressed in terms of the corresponding nucleon form factors
\cite{Marcucci:2000xy,Krebs:2016rqz}. Exchange contributions to the
axial current operator, consistent with the nuclear forces
derived in
Refs.~\cite{Epelbaum:1998ka,Epelbaum:1999dj,Epelbaum:2002gb,Epelbaum:2005fd,Epelbaum:2005bjv,Epelbaum:2007us,Bernard:2007sp,Bernard:2011zr,Krebs:2012yv,Krebs:2013kha,Springer:2025ojd}, have been calculated to the leading one-loop
order in
Ref.~\cite{Krebs:2016rqz} using the method of unitary transformation and employing
dimensional regularization to treat divergent loop integrals. As an
important consistency check, the axial current operators derived in
that paper were shown to fulfill the corresponding
continuity equation. Exchange axial currents have also
been studied using time-ordered perturbation theory \cite{Baroni:2015uza,Baroni:2016xll}, leading
to expressions that differ from those obtained
in Ref.~\cite{Krebs:2016rqz}\footnote{Nuclear forces and currents are not
  directly observable and feature intrinsic unitary ambiguities, which reflect
  their ambiguous off-shell behavior, see Ref.~\cite{Epelbaum:2025aan,Springer:2025ojd} for a recent discussion. However, it is shown in Ref.~\cite{Krebs:2020rms}
  that the box-diagram contributions to the two-nucleon axial current operator
  obtained in Refs.~\cite{Baroni:2015uza,Baroni:2016xll} and \cite{Krebs:2016rqz} are not unitary equivalent.}, see Refs.~\cite{Krebs:2020pii} for details. 
It is also important to emphasize that {\it consistently regularized}
expressions for the exchange current operators are currently only
available for the dominant tree-level contributions, which limits the
accuracy of applications to weak nuclear processes to N$^2$LO. Here,
the main complication arises from mixing dimensional
regularization in the derivation of nuclear interactions with cutoff
regularization in the nuclear Schr\"odinger equation, which is shown
in Refs.~\cite{Epelbaum:2019kcf,Krebs:2019uvm} to violate chiral symmetry. Thus, going beyond the
N$^2$LO accuracy level requires a rederivation of the
loop corrections to the exchange nuclear currents and three-nucleon
forces using a symmetry-preserving cutoff regulator instead of
dimensional regularization. Work along these lines is in progress
within the symmetry-preserving gradient-flow formulation of chiral EFT
\cite{Krebs:2023gge} and employing the path-integral approach for deriving nuclear
interactions from the effective Lagrangian \cite{Krebs:2023ljo}. 

The focus of this work is on tritium $\beta$ decay, the simplest and
one of the most extensively studied nuclear weak
processes. Traditionally, tritium $\beta$ decay is often used to calibrate
the strength of the short-range component of the dominant exchange
axial current governed by the LEC $D$. While this LEC can also be determined from strong-interaction
observables through its appearance in the 3NF, the usage of tritium $\beta$ decay was argued to be
particularly attractive due to the fact that it is not correlated with 
other low-energy observables often used to fix the 3NF, such
as the binding energies and charge radii of $A=3,4$ nuclei and
the nucleon-deuteron spin-$1/2$ S-wave scattering length \cite{Gazit:2008ma,Wesolowski:2021cni}.
The resulting parameter-free expressions for the
two-nucleon weak current operators have been applied
to make predictions
for $\beta$-decay transitions in heavier nuclei \cite{Gysbers:2019uyb,Wang:2025swg}, muon capture
reactions \cite{Acharya:2018qzk,Ceccarelli:2022cpz,Gnech:2023mvb},
the solar proton fusion and HEP processes \cite{Park:2002yp} as well as neutrino-induced
reactions \cite{Acharya:2019fij}, see also Ref.~\cite{Baroni:2018fdn} for a
related discussion. In this paper, we follow a different strategy by
using the LENPIC fitting protocol \cite{LENPIC:2018ewt,Maris:2020qne,LENPIC:2022cyu} to determine the value of the
LEC $D$ in three-nucleon scattering. We demonstrate that
the differential cross section of elastic nucleon-deuteron scattering
at intermediate energies, along with the $^3$H binding energy, provide
independent constraints on the LECs $D$ and $E$ entering the 3NF. This
allows us to make parameter-free predictions for tritium $\beta$
decay at the N$^2$LO accuracy level. Surprisingly, we find a considerable
overbinding of the GT matrix element as compared to its empirical
value, and our conclusions remain robust after performing various
consistency checks. We discuss the origin of the observed discrepancy
and implications of our findings for {\it ab initio} calculations of electroweak
reactions in chiral EFT. 

Our paper is organized as follows. In sec.~\ref{Sec:HalfLife}, we
specify the relationship between the tritium half-life
and the corresponding Fermi and Gamow-Teller matrix elements and provide
the empirical value for the GT matrix element. Next,
sec.~\ref{sec:currents} describes the employed expressions for the
single- and two-nucleon current operators. The
actual calculation of the GT matrix elements is described in detail in
sec.~\ref{Sec:performance}. Our predictions for the GT matrix element
using the LENPIC two- and three-nucleon interactions \cite{Reinert:2017usi,Maris:2020qne,LENPIC:2022cyu},
including a variety of consistency checks, are presented in
sec.~\ref{Sec:results}. The main results of this study are summarized
in sec.~\ref{Sec:summary}, where we also comment on the implications
of the observed discrepancy for chiral EFT studies of electroweak
nuclear reactions and discuss the possible next steps for resolving
the puzzle.

\section{Tritium half-life}
\label{Sec:HalfLife}
\setcounter{equation}{0}
\def\theequation{\arabic{section}.\arabic{equation}}

The rate of tritium $\beta$ decay, $^3{\rm H} \to {} ^3{\rm He} + e^- +
\bar{\nu}_e$,  can be obtained by calculating the corresponding
invariant amplitude, which depends on the nuclear matrix
element of the weak current $J^{\mu, a} = V^{\mu, a} + A^{\mu, a}$. Here,  $V^{\mu, a} \equiv (V^{0, a}, \, {\bm V}^a)$ and
$A^{\mu, a} \equiv (A^{0, a}, \, {\bm A}^a)$ refer to the vector and axial (vector)
current operators, while $a$ is the isospin index. Performing the phase space integration and taking into
account radiative corrections, one arrives at the standard and commonly used
expression for the tritium half-life \cite{Raman:1978qta}
\beq
\label{HalfLife}
(1 + \delta_R ) t f_V = \frac{K/G_V^2}{\langle \fet  F \rangle^2 +
f_A/f_V g_A^2 \langle \fet{GT} \rangle^2 }\,,
\eeq
where $\delta_R = 1.9\%$ is the radiative correction \cite{Raman:1978qta},
the Fermi functions $f_{V,A}$ have the values $f_V = 2.8355 \times
10^{-6}$ and $f_A = 2.8505 \times 10^{-6}$ \cite{Simpson:1987zz}, while the
experimental value for $K/G_V^2$ reads $K/G_V^2 = 6144.5 \pm 1.9$~s
\cite{Towner:2014uta}. Further, $\langle \fet F \rangle$ and $ \langle
\fet{GT} \rangle$ refer to the reduced nuclear matrix elements of the vector
charge and axial current operators,
respectively.
In the limit of the vanishing momentum transfer, these operators
reduce to the isospin rising Fermi  (F) and Gamow-Teller operators.
A relativistic treatment of tritium
$\beta$ decay beyond the expression in Eq.~(\ref{HalfLife})
can be found in Refs.~\cite{Simkovic:2007yi,Cavoto:2022xwo}. We, however,
emphasize that
the approximate expression in Eq.~(\ref{HalfLife}) is 
more than sufficient at the accuracy level of our considerations. 

The value of the Fermi matrix element is known to be very
close to $1$ owing to the isospin symmetry. For example, in Ref.~\cite{Baroni:2016xll}, the value
of  $\langle \fet F \rangle = 0.9998$ was obtained based on the Argonne V18 nucleon-nucleon
(NN) potential \cite{Wiringa:1994wb}, accompanied by the Urbana-IX
three-nucleon (3N) force \cite{Pudliner:1995wk}.
The deviation from $\langle \fet F
\rangle \simeq 1$ using the semilocal momentum-space-regularized  (SMS) chiral EFT
interactions from Refs.~\cite{Reinert:2017usi,Maris:2020qne,LENPIC:2022cyu} employed in this work also appears to
be very small. Specifically, using the SMS NN potentials at
N$^4$LO$^+$ along with the N$^2$LO 3NF as done in Ref.~\cite{LENPIC:2022cyu}, we find
$\langle \fet F \rangle = 1.0001\ldots 1.0013$, where the spread of
values reflects the residual dependence on the cutoff varied in the
range of $\Lambda = 400 \ldots 550$~MeV. 
Using the values for various constants in
Eq.~(\ref{HalfLife}) specified above, along with the experimental data
$(1 + \delta_R ) t f_V = 1134.6 \pm 3.1$~s \cite{Simpson:1987zz} and $g_A = 1.2756 \pm
0.0013$ \cite{Zyla:2020zbs,Gorchtein:2021fce}, we extract the empirical value    
for the GT matrix element:
\beq
{\rm GT}_{\rm emp} \equiv \langle \fet{GT} \rangle_{\rm emp} /\sqrt{3}= 0.9484 \pm 0.0019 \,,
\eeq
where the difference to the central value of ${\rm GT}_{\rm emp}  = 0.9511 \pm 0.0013$ 
from Ref.~\cite{Baroni:2016xll} is mainly due to the usage of the updated value of $g_A$.  
The uncertainty of ${\rm GT}_{\rm emp}$  induced by the interaction dependence of the Fermi matrix
element is negligibly small.

\section{Nuclear current operators}
\label{sec:currents}
\setcounter{equation}{0}
\def\theequation{\arabic{section}.\arabic{equation}}

Nuclear forces and current operators have been extensively studied in
the framework of chiral EFT, see Refs.~\cite{Epelbaum:2008ga,Machleidt:2011zz,Krebs:2020pii} for review articles.
In the Weinberg scheme, the leading-order (LO) contributions to the nuclear force
are generated by the two-nucleon one-pion exchange potential and
derivativeless contact interactions at order $Q^0$. Here
and in what follows, $Q$ refers to the chiral EFT expansion
parameter $Q \in \big( | {\bm p} |/\Lambda_b, \; M_\pi /\Lambda_b \big)$ with
${\bm p}$, $M_\pi$ and $\Lambda_b$ denoting the typical nucleon
momenta, pion mass and the EFT breakdown scale, respectively. The next-to-leading
(NLO), next-to-next-to-leading (N$^2$LO) and
next-to-next-to-next-to-leading (N$^3$LO) contributions to the nuclear
forces refer to terms of orders $Q^2$, $Q^3$ and $Q^4$,
respectively. 

The expressions for the nuclear vector and axial current operators
have so far been
derived up to the leading one-loop order for two-body
contributions by the Bochum-Bonn group using the method of unitary
transformation \cite{Kolling:2009iq,Kolling:2011mt,Krebs:2016rqz,Krebs:2019aka} and by the JLab-Pisa group using time-ordered
perturbation theory
\cite{Pastore:2008ui,Pastore:2009is,Pastore:2011ip,Baroni:2015uza}. For pioneering
studies along these lines see Refs.~\cite{Park:1993jf,Park:1995pn}. 
For the nuclear currents, the dominant terms
are generated by single-nucleon operators and thus appear at order
$Q^{-3}$ (the momentum-conserving $\delta$-function for a spectator
nucleon counts as of order $Q^{-3}$). Accordingly, NLO, N$^2$LO and
N$^3$LO corrections to the currents refer to terms of orders
$Q^{-1}$, $Q^0$ and $Q^1$, respectively. Notice that following our usual scheme, we
count the nucleon mass $m$ as a quantity of order $m \sim
\Lambda_b^2/Q$, which leads to a suppression of relativistic
corrections compared to the counting schemes with $m \sim \Lambda_b$
used, e.g., in single-baryon
chiral perturbation theory (ChPT). The JLab-Pisa group employs in their treatment
of the current operators the ChPT counting rule  $m \sim \Lambda_b$.
This results in a different hierarchy of contributions to
the current operators, with terms at orders $Q^{-2}$, $Q^{-1}$, $Q^0$
and $Q^1$ representing the NLO, N$^2$LO, N$^3$LO and N$^4$LO
corrections. 


As already mentioned in the Introduction, 
the leading loop contributions to the exchange charge and current operators of
Refs.~\cite{Kolling:2009iq,Kolling:2011mt,Krebs:2016rqz,Krebs:2019aka} 
are calculated using dimensional regularization. To prevent the appearance of
divergences when calculating the expectation values of the
current operators using nuclear wave functions, an additional cutoff
regularization needs to be employed. However, a simultaneous usage of
dimensional and cutoff regularizations  leads to a violation
of chiral symmetry and is thus inconsistent \cite{Krebs:2019uvm}. Consistently
regularized nuclear current operators beyond tree level must
be rederived using a symmetry preserving cutoff
regularization. The required methodology has been developed
in Refs.~\cite{Krebs:2023ljo,Krebs:2023gge}, but it has not yet been
applied to nuclear current operators. In this study we, therefore, restrict
ourselves to the tree-level contributions to the exchange current
operators, which appear at N$^2$LO. 

In the following subsections, we specify the expressions for the
isovector vector and axial current operators relevant for our study. We employ the nonrelativistic normalization with
$\langle {\bm p}'|{\bm p} \rangle = \delta ( {\bm p}' -  {\bm p})$ for the
nucleon states and consider the momentum-space matrix elements of 
the one- and two-nucleon currents  
$J^{\mu, a}_{\rm 1N}$  and $J^{\mu, a}_{\rm 2N}$, defined from the
corresponding operators  $\mathcal{J}^{\mu, a}_{\rm 1N}$  and $\mathcal{J}^{\mu,
  a}_{\rm 2N}$ via
\beqa
\langle {\bm p}'|  \mathcal{J}^{\mu, a}_{\rm 1N} |  {\bm p} \rangle &=:&
\delta ( {\bm p}' -  {\bm p} - {\bm k}) \, J^{\mu, a}_{\rm 1N}, \nn
\langle {\bm p}_1'  {\bm p}_1' |  \mathcal{J}^{\mu, a}_{\rm 2N} |  {\bm p}_1  {\bm p}_2\rangle &=:&
(2 \pi)^{-3} \, \delta ( {\bm p}_1' +  {\bm p}_2' -   {\bm p}_1 -
{\bm p}_2 - {\bm k}  )\,  J^{\mu, a}_{\rm 2N}\,,
\eeqa
where $a$ is an isospin index. Here, $ {\bm p}$ and  ${\bm p}'$ are
the initial and final momenta of the nucleon, while ${\bm k}$ denotes the
momentum transferred by the $W$-boson. In the second equation, the 
subscripts $i$ of the momenta ${\bm p}_i$, $ {\bm p}_i'$ refer to the
nucleon labels. Here and in what follows, $ J^{\mu, a}$ are to be understood as matrix elements in
momentum space while operators in the spin and isospin spaces.

\subsection{Single-nucleon current operator to N$^3$LO}
\label{sec:1Ncurrent}

Single-nucleon (1N) contributions to the vector and axial current operators
can be expressed in terms of the electromagnetic form factors of the
nucleon. For the vector charge density operator, the expressions
up-to-and-including N$^3$LO, derived using the method of unitary transformation, have the 
form \cite{Krebs:2019aka}
\beqa
\label{EMcharge}
V^{0, a}_{\rm 1N} &=& \tau^a \bigg( G_E^v (-k^2) +
\frac{i}{4 m^2} {\bm k} \times ({\bm p}' + {\bm p}) \cdot {\bm \sigma} \, G_M^v (-k^2)
- \frac{1}{8 m^2} \big[i  {\bm k} \times ({\bm p}' + {\bm p}) \cdot
{\bm \sigma} - k^2 \big]  G_E^v (-k^2)\bigg) ,
\eeqa
where ${\bm \tau}$ are the isospin Pauli matrices and $G_E^v (-k^2)$ and $G_M^v (-k^2)$ are the isovector
combinations of the Sachs electromagnetic form factors of the proton
and neutron,
\beq
G_E^v (-k^2) = \frac{1}{2} \big(G_E^p (-k^2) + G_E^n (-k^2) \big) ,
\qquad
G_M^v (-k^2) = \frac{1}{2} \big(G_M^p (-k^2) + G_M^n (-k^2) \big)\,.
\eeq
The energy release in the tritium $\beta$-decay process is given by
the Q-value $\Delta = M_{\rm ^3H} - M_{\rm ^3He} - m_e \simeq 0.0186$~MeV
\cite{Myers:2015lca}. Accordingly, the typical nuclear momentum
transfer
is 
$| {\bm k}| \sim \sqrt{| k_\mu k ^\mu |} \sim \sqrt{m_e \Delta} \simeq
0.138$~MeV. Therefore, to a very good approximation, one can set ${\bm k} =
0$ and the expression in Eq.~(\ref{EMcharge}) simplifies to
\beq
\label{LOVectorCurrent}
V^{0, a}_{\rm 1N} \Big|_{k^2 \; =\;  {\bm k}^2 = 0} = \frac{\tau^a}{2}\,.
\eeq

Single-nucleon contributions to the isovector axial current, derived
using the method of unitary transformation, are given
by \cite{Krebs:2016rqz}
\beqa
\label{AxialCurrent1N}
{\bm A}^{a}_{\rm 1N} &=&\tau^a \bigg[ - \frac{1}{2}  {\bm
  \sigma}  \, G_A (-k^2) + \frac{1}{8 m^2} {\bm k} {\bm k} \cdot {\bm
  \sigma} \, G_P (-k^2) \bigg] \nonumber \\
&-& \tau^a \frac{g_A k_0}{8 m} \frac{{\bm k}}{{\bm k}^2
+ M_\pi^2}\bigg[ (1 + 2 \bar \beta_9) ({\bm p}' + {\bm p} )\cdot
{\bm \sigma}
- 
 (1 + 2 \bar \beta_8) {\bm k} \cdot {\bm \sigma} \frac{{\bm p}'{}^2-
   {\bm p}^2}{{\bm k}^2 + M_\pi^2}\bigg]\nonumber \\
 &+&
 \tau^a \frac{g_A}{16m^2} \bigg[ {\bm k}  {\bm k} \cdot {\bm \sigma}
 (1 - 2 \bar \beta_8) \frac{({\bm p}'{}^2- {\bm p}^2)^2}{({\bm k}^2 +
   M_\pi^2)^2} - 2 {\bm k}
 \frac{({\bm p}'{}^2+ {\bm p}^2) {\bm k}
   \cdot {\bm \sigma}- \bar \beta_9 ({\bm p}'{}^2- {\bm p}^2)  ({\bf
     p}'+ {\bm p}) \cdot {\bm \sigma}  }{{\bm k}^2 + M_\pi^2}
 \nonumber \\
 && \hspace{1.34cm} {} + i {\bm k} \times ({\bm p}'+ {\bm p}) + {\bm k}
 {\bm k} \cdot {\bm \sigma} - ({\bm p}'+ {\bm p})  ({\bm p}'+ {\bm p})
 \cdot {\bm \sigma} + {\bm \sigma} \big( 2  ({\bm p}'{}^2+ {\bm p}^2)
 - {\bm k}^2 \big) \bigg]\,,
\eeqa
where $M_\pi$ is the pion mass, $G_A (-k^2)$ and $G_P (- k^2)$ refer to the axial and induced
pseudoscalar form factors of the nucleon, respectively, while $\bar \beta_{8}$ and  $\bar \beta_{9}$ are arbitrary phases that
parametrize the unitary ambiguity of the leading relativistic
corrections to the long-range nuclear interactions. In Refs.~\cite{Reinert:2017usi,Reinert:2020mcu},
the choice $\bar \beta_8 = - \bar \beta_9 = 1/4$ was employed, which
corresponds to the minimal nonlocality of the $1/m^2$-corrections to
the one-pion exchange NN force. Again, for the kinematics of the
tritium $\beta$ decay, the expression for the axial current simplifies
to
\beq
\label{AxialCurrent1NFinal}
{\bm A}^{a}_{\rm 1N} \Big|_{k^2 \; =\;  {\bm k}^2 = 0} =
-\frac{g_A}{2} \tau^a {\bm \sigma}
+ \frac{g_A}{16 m^2}  \tau^a \big[ 2  ({\bm p}'{}^2+ {\bm p}^2)  {\bm
  \sigma} - 
 ({\bm p}'+ {\bm p})  ({\bm p}'+ {\bm p})
 \cdot {\bm \sigma} \big]
\,,
\eeq
where we have used $G_A (0) = g_A$.  Here, the first term is the LO
contribution, while the relativistic corrections start contributing at N$^3$LO
in our power counting scheme.

\subsection{Two-nucleon current at N$^2$LO} 

The first contributions to the isovector vector two-nucleon charge density
operator $V^{0, a}_{\rm 2N}$ appear at N$^3$LO with the corresponding
expressions given in
Refs.~\cite{Kolling:2009iq,Kolling:2011mt}, see also
Ref.~\cite{Krebs:2019aka}. As already mentioned, the corresponding
loop contributions need to be
rederived using the symmetry-preserving  gradient flow regularization \cite{Krebs:2023ljo,Krebs:2023gge},
but they are beyond the accuracy level of this study. In contrast, the
leading 
two-body contributions to the axial current stem from
tree-order diagrams already at N$^2$LO. The corresponding unregularized
expressions have the form
\beqa
\label{CurrentTree2}
 {\bm A}_{{\rm 2N}}^{a} &=& \frac{g_A}{2 F_\pi^2}
\frac{{\bm \sigma}_1 \cdot {\bm q}_1}{{\bm q}_1^2 + M_\pi^2} \bigg\{
\tau_1^a 
\bigg[ - 4 c_1 M_\pi^2 \frac{\bm k}{{\bm k}^2 + M_\pi^2}
+ 2 c_3 \bigg( {\bm q}_1 - \frac{{\bm k} \, {\bm  k} \cdot {\bf
    q}_1}{{\bm k}^2 +
  M_\pi^2} \bigg) \bigg]\\
&+&c_4 [ {\bm \tau}_1 \times {\bm \tau}_2 ]^a \bigg({\bm q}_1 \times
{\bm \sigma}_2 - \frac{{\bm k} \, {\bm k} \cdot {\bm q}_1 \times 
 {\bm \sigma}_2}{{\bm k}^2 + M_\pi^2} \bigg)-\frac{\kappa_v}{4 m}  [
{\bm \tau}_1 \times {\bm \tau}_2 ]^a
{\bm k}\times{\bm \sigma}_2\bigg\}  
-\frac{1}{4} D\,
\tau_1^a \bigg({\bm \sigma}_1-\frac{{\bm k}\, {\bm \sigma}_1\cdot{\bm
    k}}{{\bm k}^2+M_\pi^2}\bigg)
+ \; 1 \leftrightarrow 2\,, 
\nonumber
\eeqa
where $1 \leftrightarrow 2$  refers to the contribution obtained by
interchanging the nucleon labels,
${\bm \sigma}_i$ and ${\bm \tau}_i$ are the spin and
isospin Pauli matrices of the nucleon $i$, while $\fet q_{i} = \fet p_i  ' - \fet p_i$ denote the momentum transfer of the nucleon $i$.
Further  $F_\pi$ is the pion decay constant, $c_i$ and $D$ are further 
LECs, while
$\kappa_v = 3.706$ [n.m.] is the isovector anomalous magnetic moment of the nucleon.  

Semilocal momentum-space regularization of the exchange currents at
tree level,
consistent with the chiral EFT
two- and three-body forces of Refs.~\cite{Reinert:2017usi} and
\cite{Maris:2020qne}, respectively, can be implemented via
\beqa
\label{Regularized}
{\bm A}_{{\rm 2N, \; reg.}}^{a} &=& \frac{g_A}{2 F_\pi^2}
\frac{{\bm \sigma}_1 \cdot {\bm q}_1}{{\bm q}_1^2 + M_\pi^2} \,
e^{-\frac{{\bm q}_1^2 + M_\pi^2}{\Lambda^2}}
\bigg\{ 
\tau_1^a \bigg[ - 4 c_1 M_\pi^2 \frac{\bm k}{{\bm k}^2 + M_\pi^2}
+ 2 c_3 \bigg( {\bm q}_1 - \frac{{\bm k} \, {\bm k} \cdot {\bf
    q}_1}{{\bm k}^2 +
  M_\pi^2} \bigg) \bigg]\notag\\
&+& c_4 [ {\bm \tau}_1 \times {\bm \tau}_2 ]^a \bigg({\bm q}_1 \times
{\bm \sigma}_2 - \frac{{\bm k} \, {\bm  k} \cdot {\bm q}_1 \times 
 {\bm \sigma}_2}{{\bm k}^2 + M_\pi^2} \bigg)-\frac{\kappa_v}{4 m}  [
{\bm \tau}_1 \times  {\bm \tau}_2 ]^a
{\bm k}\times{\bm \sigma}_2\bigg\}
- \; \frac{1}{4} D\,
\tau_1^a \bigg({\bm \sigma}_1-\frac{{\bm k}\, {\bm \sigma}_1\cdot {\bf
    k}}{{\bm k}^2+M_\pi^2}\bigg)
\, e^{-\frac{{\bm p}^2  + {{\bm p}^\prime}^2}{\Lambda^2}}\notag\\
&+& \frac{g_A C}{2 F_\pi^2} \, e^{-\frac{{\bm q}_1^2 + M_\pi^2}{\Lambda^2}}
\bigg\{
2 c_3 \tau_1^a \bigg( {\bm \sigma}_1 - \frac{{\bm  k}  \, 
  {\bm \sigma}_1 \cdot {\bm  k}}{{\bm k}^2 + M_\pi^2} \bigg) + c_4
[{\bm \tau}_1
\times  {\bm \tau}_2 ]^a \bigg( {\bm \sigma}_1 \times {\bm  \sigma}_2
- \frac{{\bm  k}  \, {\bm \sigma}_1 \times {\bm  \sigma}_2
 \cdot {\bm  k}}{{\bm k}^2 + M_\pi^2} \bigg) 
\bigg\} \;
+ \; 1 \leftrightarrow 2\,,
\eeqa
where $\Lambda$ is the cutoff, while ${\bm p} = \frac{1}{2} ({\bm  p}_1 - {\bm  p}_2 )$ and ${\bm p}
' = \frac{1}{2} ({\bm  p}_1 ' - {\bm  p}_2  ' )$. Furthermore, $C$ is
a subtraction constant given by 
\beq
C = -\frac{\Lambda \left(\Lambda ^2-2 M_\pi^2 \right) + 2 \sqrt{\pi } M_\pi^3 e^{\frac{M_\pi^2}{\Lambda ^2}}
   \text{erfc}\left(\frac{M_\pi}{\Lambda }\right)}{3 \Lambda ^3} \,,
 \label{BIGC}
 \eeq
 where $\text{erfc}(x)$ denotes the complementary error function.
The subtraction terms represent a {\it convention}, which ensures that
the regularized long-range contributions to the
axial current $\propto c_{3,4}$ vanish in coordinate space at the relative distance
between the nucleons ${\bm  r}_{12} = 0$.  The same convention is
employed for 3NFs, see appendix \ref{appen:3NF} for
details. 
The pion-pole contributions in Eq.~(\ref{Regularized}) show a clear
correspondence to the regularized expression for the leading
three-nucleon force in Eq.~(\ref{3NFN2LO}), as discussed in
detail in Ref.~\cite{Krebs:2016rqz}.  Dropping the contributions
proportional to the momentum of the weak source ${\bm k}$, which are
heavily suppressed for tritium $\beta$ decay, we end up with the
simplified expression
\beqa
\label{RegularizedFinal}
{\bm A}_{{\rm 2N, \; reg.}}^{a} &=& \frac{g_A}{2 F_\pi^2}
\frac{{\bm \sigma}_1 \cdot {\bm q}_1}{{\bm q}_1^2 + M_\pi^2} \,
e^{-\frac{{\bm q}_1^2 + M_\pi^2}{\Lambda^2}}
\big( 
2 c_3 \tau_1^a  {\bm q}_1 
+ c_4 [ {\bm \tau}_1 \times {\bm \tau}_2 ]^a  {\bm q}_1 \times
{\bm \sigma}_2
\big)
- \; \frac{1}{4} D\,
\tau_1^a   {\bm \sigma}_1
\, e^{-\frac{{\bm p}^2  + {{\bm p}^\prime}^2}{\Lambda^2}}\notag\\
&+& \frac{g_A C}{2 F_\pi^2} \, e^{-\frac{{\bm q}_1^2 + M_\pi^2}{\Lambda^2}}
\big(
2 c_3 \tau_1^a  {\bm \sigma}_1 + c_4
[{\bm \tau}_1
\times  {\bm \tau}_2 ]^a {\bm \sigma}_1 \times {\bm  \sigma}_2
\big)\;
+ \; 1 \leftrightarrow 2
\eeqa
to be used in the course of the present study. We further emphasize that
gradient-flow regularized expressions for the exchange axial current, which can
be derived using the methods presented in Ref.~\cite{Krebs:2023gge}, are expected to differ
from the SMS expression in Eq.~(\ref{Regularized}). However, for the kinematics with $k^\mu = 0$
relevant for this study, no differences are expected and the
expression in Eq.~(\ref{RegularizedFinal}) should remain valid.

\section{Technical performance} \label{Sec:performance}
\setcounter{equation}{0}
\def\theequation{\arabic{section}.\arabic{equation}}

To compute the tritium half-life, one needs to calculate the
expectation values of the weak current in the $^3$H and $^3$He states. 
In this section, we provide details on the numerical evaluation of the
required matrix elements.

\subsection{The $^3$H and $^3$He wave functions}


Three-nucleon states are represented using a momentum space partial-wave basis
\begin{eqnarray}  
| \Psi  \rangle  =
\sum\limits_{{\alpha}} \,
\int dp p^2 \int dq q^2 \, 
| p q {\alpha}  \rangle \, \phi_{{\alpha}} ( p , q) \, .
\label{pwdj13}
\end{eqnarray}  
The construction of the $|p q {\alpha}_b  \rangle  $  basis states
starts with the two-nucleon states
$ | p \alpha_2  \rangle \, \equiv \, | p (l s ) j m_j ; t m_t  \rangle$,
where $p$ represents the magnitude of the relative momentum,
$l$ and $p$ are the relative angular momentum and spin, respectively,
while $j$ is the total angular
momentum with the corresponding magnetic quantum number $m_j$.
Since nucleons are treated as identical particles, this set of quantum numbers is supplemented by the 2N isospin $t$ and its projection $m_t$.
The final 3N states 
\beq
| p q \alpha  \rangle \, \equiv \, |  p (l s ) j \, q (\lambda \frac12 )I 
\, (j I)J m_J ; T m_T  \rangle 
\eeq
are built upon the subsystem $(2,3)$ quantum numbers 
and carry additional information about 
nucleon 1: the magnitude $q$ of its relative momentum with 
respect to the center-of-mass of the $(2,3)$-subsystem,
the relative orbital angular momentum $\lambda$, total angular
momentum $I$ of nucleon 1 
and, finally, about the total 3N angular
momentum $J$ with the magnetic quantum number $m_J$, stemming from coupling 
of the angular momenta  $j$ and  $I$. 
The total 3N isospin state, 
$ | (t \frac12 ) T m_T  \rangle$
is obtained by coupling the isospin $t$ of the $(2,3)$ subsystem and
the isospin 
$1/2$ of the spectator nucleon. By construction, such complete 3N partial-wave states are
antisymmetrized in the $(2,3)$ subsystem. More details about the
partial wave basis and the employed conventions can be found in
Ref.~\cite{physrep}. 

\subsection{Nuclear matrix elements of the weak current}

The isospin component of the weak current operator relevant for
the tritium $\beta$ decay is given by
\beq
\label{CurrentChargeRaising}
 J^{\mu,+} \equiv  J^{\mu,a=1} + i  J^{\mu,a=2}
\eeq
and is proportional to the charge-raising operators $\frac{1}{2}
\tau_i^+ \equiv \frac{1}{2} (\tau_i^{a=1} + i \tau_i^{a=2})$ and
$[{\bm \tau}_1 \times {\bm \tau}_2]^+ \equiv [{\bm \tau}_1 \times {\bm
  \tau}_2]^{a = 1} + i  [{\bm \tau}_1 \times {\bm
  \tau}_2]^{a = 2}$. Below, we describe the calculation of  the
matrix elements of $ J^{\mu,+}$ between the $^3\text{H}$ and
$^3\text{He}$ states.

\subsubsection{Treatment of the single-nucleon current}

In order to calculate the single-nucleon matrix elements 
\[
\big\langle  {^3\text{He}},  m_{^3\text{He}},  {\bm P}_f  \big|
{J}_{1\text{N},\, (1)}^{\mu,+} +
{J}_{1\text{N},\, (2)}^{\mu,+} + {J}_{1\text{N},\, (3)}^{\mu,+} \big|
 {^3\text{H}},  m_{^3\text{H}},  {\bm P}_i  \big\rangle,
\]
where both the spin $J_b$ of $^3$H and the spin $J$ of $^3$He are
assumed to be $1/2$, $m_{^3\text{H}}$ and $m_{^3\text{He}}$ denote
the spin projections of $^3$H and $^3$He, while ${\bf
  P}_i$, ${\bm P}_f$ are, respectively, the total momenta of $^3$H and $^3$He, we utilize the symmetry of the nuclear states and calculate only the contribution from nucleon 1 in a well known form~\cite{physrep}:
\begin{multline}
\big\langle  {^3\text{He}},  m_{^3\text{He}},  {\bm P}_f  \big|  {J}_{1\text{N},\, (1)}^{\mu,+} +
{J}_{1\text{N},\, (2)}^{\mu,+} + {J}_{1\text{N},\, (3)}^{\mu,+} \big|
 {^3\text{H}},  m_{^3\text{H}},  {\bm P}_i  \big\rangle\\
 = 
3\, \sum\limits_{\alpha} \int dp p^2 \int dq q^2 \,
\big\langle  {^3\text{He}},  m_{^3\text{He}},  {{\bm P}_f} \big|  p q \alpha , {{\bm P}_f} \big\rangle \, 
\big\langle  p q \alpha , {{\bm P}_f} 
\big| 
{J}_{1\text{N},\, (1)}^{\mu,+}
\big|
 {^3\text{H}},  m_{^3\text{He}},  {{\bm P}_i}  \big\rangle \, ,
\end{multline}
where a general formula for the crucial matrix element 
$
\big\langle  p q \alpha , {{\bm P}_f} 
\big| 
{J}_{1\text{N},\, (1)}^{\mu,+}
\big|
 {^3\text{H}}, m_{^3\text{H}} , {{\bm P}_i} \big\rangle
$
follows essentially  from the definition of the 3N partial wave
states. While we do allow for the total 3N isospin-$T= 3/2$ admixtures
in the nuclear states, the projections of the total 3N isospin are
fixed to $m_{T_b} = -1/2$ for the $^3$H nucleus and $m_T = 1/2$ for
the final $^3$He state: 
\begin{align}  
&\big\langle p q \alpha , {\bm P}_f \big| {J}_{1\text{N},\, (1)}^{\mu,+} \big| {^3\text{H}}, m_{^3\text{H}} , {\bm P}_i  \big\rangle 
\notag \\
&\hspace{1.0cm}
	=\sum\limits_{{\alpha}_b}
\delta_{ l ,  l_b}  
\delta_{ s , s_b} 
\delta_{ j , j_b} \; 
	\bigg\langle  \bigg( t \frac12 \bigg) T \, m_T \bigg| 
	 \frac{\tau_1^+}{2}
	\bigg| 
 \bigg( t_b \frac12 \bigg) T_b \, m_{T_b}  \bigg\rangle 
\;
  \sum\limits_{m_j}  
c ( j,  I, J ; m_j , m_J - m_j , m_J ) \,
\notag \\
&\hspace{1.0cm} \times 
c \bigg( j_b,  I_b, \frac12 ; m_j , m_{^3\text{H}} - m_j , m_{^3\text{H}}   \bigg) \,
     \sum\limits_{m_\lambda, m_{\lambda_b}}  
c \bigg( \lambda , \frac12 , I ; m_\lambda , m_J - m_j - m_\lambda , m_J - m_j \bigg)  
\notag \\
&\hspace{1.0cm} \times
c \bigg( \lambda_b , \frac12 , I_b ; m_{\lambda_b} , m_{^3\text{H}} - m_j  - m_{\lambda_b} , m_{^3\text{H}} - m_j \bigg) \,
            \int d \hat{\bm q} \;
Y^*_{\lambda\, m_\lambda } \big( \hat{\bm q} \big) \,
Y_{\lambda_b \, m_{\lambda_b} } \big( \hat{\bm Q}_{{\bm q}, {\bm
            P}_f,{\bm P}_i} \big) \,
\phi_{{\alpha}_b} \big( p , \big| {\bm Q}_{{\bm q}, {\bm
            P}_f,{\bm P}_i}   \big| \big)
\notag \\
&\hspace{1.0cm} \times  \bigg\langle  \frac12 \, m_J - m_j - m_\lambda
            \bigg| \bigg\langle  {\bm q} + \frac13 {\bm P}_f  
\bigg| {J}_{1\text{N},\, (1),\,\text{spin}}^{\mu} \bigg|  {\bm q} - \frac23 {\bm P}_f + {\bm P}_i  \bigg\rangle 
\bigg| \frac12 \, m_{^3\text{H}} - m_j - m_{\lambda_b}   \bigg\rangle \, ,
\label{pwdj1}
\end{align}  
where ${\bm Q}_{{\bm q}, {\bm
            P}_f,{\bm P}_i} \equiv {\bm q} - \frac{2}{3} ({\bm P}_f -
        {\bm P}_i)$, ${J}_{1\text{N},\, (1),\, \text{spin}}^{\mu}$ is
        the single-nucleon current ${J}_{1\text{N},\, (1)}^{\mu, +}$ without the isospin operator
        $\tau_1^+/2$, 
$c \left( j_1 , j_2 , j_3 ; m_1 , m_2 , m_3 \, \right) $
are Clebsch-Gordan coefficients and $\hat{\bm a}$ denotes a unit
vector ${\bm a}/|{\bm a} |$.
In the momentum-dependent spin matrix elements 
\[
 \bigg\langle  \frac12 \, m_J - m_j - m_\lambda \bigg| \bigg\langle
 {\bm q} + \frac13 {\bm P}_f  
\bigg| {J}_{1\text{N},\, (1),\, \text{spin}}^{\mu} \bigg|  {\bm q} - \frac23 {\bm P}_f + {\bm P}_i  \bigg\rangle 
\bigg| \frac12 \, m_{^3\text{H}} - m_j - m_{\lambda_b}   \bigg\rangle,
\]
the linear combinations $ {\bm q} - \frac23 {\bm P}_f + {\bm P}_i$
and
${\bm q} + \frac13 {\bm P}_f $
play the role of the initial and final nucleon momenta. 
The isospin matrix element yields
\begin{align}
    	\bigg\langle  \bigg( t \frac12 \bigg) T \, m_T \bigg| 
	 \frac{\tau_1^+}{2}
	\bigg| 
 \bigg( t_b \frac12 \bigg) T_b \, m_{T_b} \, \bigg\rangle 
 &=
 \delta_{ t , t_b} 
	\delta_{ m_T , m_{T_b}+1 } \, \sqrt{3} \, 
	(-1)^{ t + \frac12 - T_b }   \sqrt{ 2 T_b + 1} \, 
	\left\{
		\begin{array}{ccc}
			1 & \frac12 & \frac12 \\\
			t & T & T_b 
		\end{array}
		\right\} \, 
	c( 1, T_b, T ; 1, m_{T_b} , m_{T}) \, .
\end{align}
For the leading-order contributions to the single-nucleon weak
current, $J_{\rm 1N ,\, (1)}^{0, + \, \rm (LO)} = V_{\rm 1N ,\, (1)}^{0, + \, \rm
  (LO)} = \frac{1}{2} \tau_1^+$ and 
${\bm J}_{\rm 1N ,\, (1)}^{+ \, \rm (LO)} = {\bm A}_{\rm 1N ,\, (1)}^{+ \, \rm
  (LO)} = -\frac{g_A}{2} \tau_1^+ {\bm \sigma}_1$, we obtain 
a quasi-analytical result for the whole matrix element (\ref{pwdj1}),
especially when we assume that not only the initial $^3$H but also the
final $^3$He nucleus is at rest with 
${\bm P}_i =  {\bm P}_f = 0 $:
\beqa
\label{pwdENR0}
\big\langle p q \alpha , {\bm P}_f \big| 
J_{1\text{N},\, (1)}^{0,+\,{\rm (LO)}} 
\big| {^3\textrm{H}}, m_{^3\text{H}} , {\bm P}_i \big\rangle \Big|_{{\bm
  P}_f={\bm P}_i=0}  
&=&\delta_{ m_J  , m_{^3\text{H}} } \,
\delta_{ J  , \frac12 }  
	\sum\limits_{{\alpha}_b}
\, \delta_{ l  , l_b} \, 
\delta_{ s , s_b} \, 
\delta_{ j , j_b} \, 
\delta_{ \lambda  , \lambda_b} \, 
\delta_{ I  , I_b} \nonumber \\
&\times &
	\bigg\langle  \bigg( t \frac12 \bigg) T \, m_T \bigg| 
	 \frac{\tau_1^+}{2}
	\bigg| 
 \bigg( t_b \frac12 \bigg) T_b \, m_{T_b}  \bigg\rangle 
\;
\phi_{{\alpha}_b} ( p , q )\,.
\eeqa
For the spherical component $\kappa$ of the spatial current
${\bm J}_{1\text{N},\, (1)}^{+\,{\rm (LO)}}$, we find
\beqa
\big \langle p q \alpha , {\bm P}_f \big| 
{J}_{1\text{N},\, (1)}^{\kappa,+\,{\rm (LO)}}
\big| {^3\text{H}}, m_{^3\text{H}} , {\bm P}_i \big\rangle \Big|_{{\bm
  P}_f={\bm P}_i=0} &=&
\delta_{ m_J  , m_{^3\text{H}} + \kappa } \,
2 \sqrt{3} \, g_A \, 
	\sum\limits_{{\alpha}_b}
\delta_{ l  , l_b}  
\delta_{ s , s_b}  
\delta_{ j , j_b}  
\delta_{ \lambda  , \lambda_b} \; \phi_{{\alpha}_b} ( p , q)
\nonumber \\
&\times &
	\bigg\langle  \bigg( t \frac12 \bigg) T \, m_T \bigg| 
	 \frac{\tau_1^+}{2}
	\bigg| 
 \bigg( t_b \frac12 \bigg) T_b \, m_{T_b}  \bigg\rangle 
\;
(-1)^{ 1 + j + \lambda + I + I_b } \,
\sqrt{ ( 2 I + 1 ) \, ( 2 I_b + 1 ) \, } \nonumber \\[3pt]
&\times&
	\left\{
		\begin{array}{ccc}
			1 & \frac12 & \frac12 \\
			\lambda & I & I_b 
		\end{array}
		\right\} \, 
			\left\{
		\begin{array}{ccc}
			1 & I_b & I \\
			j & J & \frac12 
		\end{array}
		\right\} \,
			c\bigg( 1, \frac12, J ; \kappa , m_{^3\text{H}} , m_{J}\bigg) \, .
\label{pwdENRV}
\eeqa
For the relativistic correction to the axial current operator given by
the last term in Eq.~(\ref{AxialCurrent1NFinal}), we use the so-called automatized partial wave
decomposition method \cite{apwd1,apwd2}, i.e.~we calculate the  
momentum-dependent spin matrix elements 
$
\big\langle  \frac12 m_1'  \big| \big\langle {\bm p}_1' 
\big| {J}_{1\text{N},\, (1),\, \text{spin}}^{\mu,+} \big|
{\bm p}_1  \big\rangle \big| \frac12 m_1 \, \Big\rangle \,
$
analytically using a computer algebra system 
and perform the double integral over
$
\int d \hat{\bm q} \, \equiv \,
	\int\limits_0^{2 \pi} d \phi_q \, 
	\int\limits_0^{\pi} d \theta_q \sin \theta_q 
$
numerically.

\subsubsection{Treatment of the two-nucleon current}

For two-nucleon matrix elements 
\[
\langle  {^3\text{He}} , m_{^3\text{He}} , {{\bm P}_f}  |
J_{2\text{N}, \, (2,3)}^{\mu,+} +
J_{2\text{N}, \, (3,1)}^{\mu,+} + J_{2\text{N}, \, (1,2)}^{\mu,+} | 
 {^3\text{H}} , m_{^3\text{H}} , {{\bm P}_i}  \rangle,
\]
the same symmetry argument holds, and it is sufficient to calculate 
$
\langle  p q \alpha , {{\bm P}_f} 
| 
J_{2\text{N}, \, (2,3)}^{\mu,+}
|
 {^3\text{H}} , m_{^3\text{He}} , {{\bm P}_i}  \rangle
$.
We assume a general form of the two-body current operator
\begin{eqnarray}
  \label{2NCurrentFunctionMomenta}
 \langle {\bm p}_1'  {\bm p}_2' {\bm p}_3' |   J_{2\text{N}, \, (2,3)}^{\mu,+} |
  {\bm p}_1 {\bm p}_2 {\bm p}_3  \rangle &=& \delta (  {\bm p}_1'  - {\bm p}_1) \;
  \langle  {\bm p}_2' {\bm p}_3' |   J_{2\text{N}, \, (2,3)}^{\mu,+} |
                                             {\bm p}_2 {\bm p}_3  \rangle \nonumber \\
                                         &\equiv&
                                              \delta \bigg( {\bm q}'  - {\bm q} + \frac13 (  {\bm P}' - {\bm P}  )  \bigg) \; 
 J_{2\text{N}, \, (2,3)}^{\mu,+} ( {\bm p}'  , {\bm q}' , {\bm P}' ; \, {\bm p} , {\bm q} , {\bm P} ) \,,
\end{eqnarray}
where we have introduced  the relative Jacobi momenta 
$ {\bm p} = \frac12 ( {\bm p}_2 - {\bm p}_3 ) $,
$ {\bm q} = \frac23 \big( {\bm p}_1  - \frac12 ( {\bm p}_2  + {\bm
  p}_3) \big)$ and
the total 3N momentum 
$ {\bm P}  = {\bm p}_1 + {\bm p}_2 +{\bm p}_3 $ in the initial state, as well as the
corresponding momenta $ {\bm p}' = \frac12 ( {\bm p}_2' - {\bm p}_3' ) $,
$ {\bm q}' = \frac23 \big( {\bm p}_1'  - \frac12 ( {\bm p}_2'  + {\bm
  p}_3') \big)$ and $ {\bm P}'  = {\bm p}_1' + {\bm p}_2' +{\bm p}_3'
$ in the final state. 
Then, standard steps lead to 
\beqa
&&\langle p q \alpha , {\bm P}_f | J_{2\text{N}, \, (2,3)}^{\mu,+} | {^3\text{H}}, m_{^3\text{H}} , {\bm P}_i  \rangle  
\; = \;
\sum\limits_{{\alpha}_b}
\sum\limits_{m_j,m_{j_b}}
c ( j,  I, J ; m_j , m_J - m_j , m_J )\;
c \bigg( j_b,  I_b, \frac12 ; m_{j_b} , m_{^3\text{H}} - m_{j_b} , m_{^3\text{H}} \bigg)
\notag\\
&&\hspace{1cm}\times
\sum\limits_{m_l,m_{l_b}}  
c ( l,  s, j ; m_l , m_j - m_l , m_j )
\; c ( l_b,  s_b, j_b ; m_{l_b} ,  m_{j_b} - m_{l_b} , m_{j_b} ) 
\notag\\
&&\hspace{1cm}\times
\sum\limits_{m_\lambda,m_{\lambda_b}}
c \bigg( \lambda , \frac12 , I ; m_\lambda , m_J - m_j - m_\lambda , m_J - m_j \bigg) \, 
c \bigg( \lambda_b , \frac12 , I_b ; m_{\lambda_b} , m_{^3\text{H}} - m_{j_b}  - m_{\lambda_b} , m_{^3\text{H}} - m_{j_b} \bigg) 
\notag\\
&&\hspace{1cm}\times
\sum\limits_{m_t,m_{t_b}}  
\delta_{ m_T - m_t ,  m_{T_b} - m_{t_b}  }\, 
c \bigg( t , \frac12 , T ; m_t , m_T - m_t , m_T \bigg) \;  
c \bigg( t_b , \frac12 , T_b ; m_{t_b} , m_{T_b} - m_{t_b} , m_{T_b} \bigg) 
\notag\\
&&\hspace{1cm}\times
\int dp_b \, p_b^2 \, \int d \hat{\bm p}_b \, \int d \hat{\bm p} \,
Y^*_{l \, m_l } ( \hat{\bm p} ) \,
Y_{l_b \, m_{l_b} } ( \hat{\bm p}_b )
\int d \hat{\bm q} \,
Y^*_{\lambda\, m_\lambda } ( \hat{\bm q} ) \,
Y_{\lambda_b \, m_{\lambda_b} } \bigl(\hat{\boldsymbol{\mathcal{Q}}}_{{\bm q}, {\bm P}_f,{\bm P}_i} \big)\; 
\phi_{{\alpha}_b} \big( p_b ,\big|{\boldsymbol{\mathcal{Q}}}_{{\bm q}, {\bm P}_f,{\bm P}_i}\big|  \big)
\notag\\[4pt]
&&\hspace{1cm}\times
\big\langle t m_t \big| \big\langle s \, m_j - m_l \big|
J_{2\text{N}, \, (2,3)}^{\mu,+} \big( {\bm p}  , {\bm q} , {\bm P}_f ;
\,  {\bm p}_b , {\bm q} + \tfrac13({\bm P}_f - {\bm P}_i) , {\bm P}_i \big)
\big| s_b \, m_{j_b}-  m_{l_b} \big\rangle \big| t_b m_{t_b} \big\rangle ,
\label{pwdj23}
\eeqa  
where ${\boldsymbol{\mathcal{Q}}}_{{\bm q}, {\bm P}_f,{\bm P}_i}={\bm q} + \tfrac13({\bm P}_f - {\bm P}_i)$ and the individual momenta in the initial and final states are now given as: 
$ {\bm p}_{b,1} = {\bm q} + \tfrac13 {\bm P}_f$,
$ {\bm p}_{b,2} =  {\bm p}_b - \tfrac12 {\bm q} + \tfrac12 {\bm P}_i - \tfrac16 {\bm P}_f$,
$ {\bm p}_{b,3} =  -{\bm p}_b - \tfrac12 {\bm q} + \tfrac12 {\bm P}_i - \tfrac16 {\bm P}_f$,
$ {\bm p}_1 = {\bm q} + \tfrac13 {\bm P}_f$,
$ {\bm p}_2 = {\bm p} - \tfrac12 {\bm q} + \tfrac13 {\bm P}_f$ and
$ {\bm p}_3 = -{\bm p} - \tfrac12 {\bm q} + \tfrac13 {\bm P}_f$.

An additional simplification emerges if we assume $ {\bm P}_i = {\bm
  P}_f = 0 $ and make use of the fact that in this case and for the current operator specified in Eq.~(\ref{Regularized}), the function
$J_{2\text{N}, \, (2,3)}^{\mu,+} ( {\bm p}'  , {\bm q}' , {\bm P}' ;
\, {\bm p} , {\bm q} , {\bm P} )$, introduced in
Eq.~(\ref{2NCurrentFunctionMomenta}), reduces to 
just  $J_{2\text{N}, \, (2,3)}^{\mu,+} ( {\bm p}' ; 
\, {\bm p} )$. Accordingly, the integration over $\hat{\bm q}$
separates out and can be trivially performed using the orthogonality of
the spherical harmonics, leading to our final result 
\beqa
\label{pwdj23.2}
&&\langle p q \alpha , {\bm P}_f | J_{2\text{N}, \, (2,3)}^{\mu,+} |
{^3\textrm{H}}, m_{^3\text{H}} , {\bm P}_i  \rangle  \Big|_{{\bm
  P}_f={\bm P}_i=0}  
\; = \;
\sum\limits_{{\alpha}_b}\delta_{ \lambda , \lambda_b } \delta_{ I , I_b   } 
\sum\limits_{m_j,m_{j_b}}  
c ( j,  I, J ; m_j , m_J - m_j , m_J )
\notag\\
&&\hspace{1cm}\times\, 
c \bigg( j_b,  I_b, \frac12 ; m_{j_b} , m_{^3\text{H}} - m_{j_b} , m_{^3\text{H}} \bigg)
\sum\limits_{m_t,m_{t_b}}  
\delta_{ m_T - m_t ,  m_{T_b} - m_{t_b}  }
\, c \bigg( t , \frac12 , T ; m_t , m_T - m_t , m_T\bigg) 
\\
&&\hspace{1cm}\times\, 
c \left( t_b , \frac12 , T_b ; m_{t_b} , m_{T_b} - m_{t_b} , m_{T_b} \,  \right) \,
\int dp_b \, p_b^2 \, 
\phi_{{\alpha}_b} ( p_b , q ) 
\, \int d \hat{\bm p} \, \int d \hat{\bm p}_b \,
\sum\limits_{m_l,m_{l_b}}  
c ( l,  s, j ; m_l , m_j - m_l , m_j )
\notag\\
&&\hspace{1cm}\times \, 
c ( l_b,  s_b, j_b ; m_{l_b} ,  m_{j_b} - m_{l_b} , m_{j_b} ) \,
Y^*_{l \, m_l } ( \hat{\bm p} ) \,
Y_{l_b \, m_{l_b} } ( \hat{\bm p}_b )\; 
\big\langle t m_t \big| \big\langle s \, m_j - m_l \big|
J_{2\text{N}, \, (2,3)}^{\mu,+} ( {\bm p}' ; 
\, {\bm p} )
 \big| s_b \, m_{j_b}-  m_{l_b} \big\rangle \big| t_b m_{t_b} \big\rangle .
\nonumber
\eeqa  
When the summations over $m_t$ and $m_{t_b}$ are evaluated, only four
isospin combinations yield non-zero contributions for the
isospin-raising operators present in Eq.~(\ref{CurrentChargeRaising}),
namely $\{t, m_t, 
t_b, m_{t_b}\} = \{ 0, 0, 1, -1\}$, $\{ 1, 0, 1, -1\}$, $\{ 1, 1, 1,
0\}$ and $\{ 1, 1, 0, 0\}$. 
Additionally, the second and third cases lead to the same result since 
for $t = t_b = 1$, the operator $ [\fet \tau_2 \times \fet \tau_3 ]^+ $ gives no contribution and $ \langle \left( \frac12 , \frac12 \right) 1 0 \left|  \frac12  \tau_2^+  \right|  \left( \frac12 , \frac12 \right) 1 -1 \rangle 
=
 \langle \left( \frac12 , \frac12 \right) 1 1 \left|  \frac12\tau_2^+  \right|  \left( \frac12 , \frac12 \right) 1 0 \rangle 
$.
The momentum-dependent spin-isospin matrix elements 
$ \langle s \, m_j - m_l | J_{2\text{N}, \, (2,3)}^{\mu,+} ( {\bm p}' ; 
\, {\bm p} )
 | s_b \, m_{j_b}-  m_{l_b} \rangle  $
are calculated analytically  using
a computer algebra system and constitute the essential part of the
Fortran code, which  
evaluates the fourfold integrals over $\int d \hat{\bm p} $ and $\int
d \hat{\bm p}_b$ numerically. 

The 3N bound-state wave functions take into account  all 3N partial waves with 
$j_b \le 5$. In the calculation of the single-nucleon matrix elements,
all these channels are taken into account. The partial wave decomposition 
of the 2N axial current operators has been carried out by taking into
account all 2N partial wave states with $ j \le 2$. 

In principle, in order to study the GT part of the tritium $\beta$-decay rate, it is sufficient to calculate matrix elements of only one vector component, since for $ {\bm P}_i= {\bm P}_f= 0$ the three results
($\kappa = 0, \pm 1$)
\[
\frac12 \sum\limits_{m_{^3\text{He}} , m_{^3\text{H}}  } \left|
\big\langle \, {^3\text{He}} , m_{^3\text{H}} , {\bm P}_f  \big| 
\big( J_{1\text{N}, \, (1)}^{\kappa,+} +J_{2\text{N}, \, (2,3)}^{\kappa,+} \big) \big| 
 {^3\text{H}} , m_{^3\text{H}} , {\bm P}_i \big\rangle \Big|_{{\bm
  P}_f={\bm P}_i=0}  
 \right|^2 
\]
should be identical. This property is used as an additional accuracy
test of our calculations. 

\section{N$^2$LO  predictions for the Gamow-Teller reduced matrix element} \label{Sec:results}
\setcounter{equation}{0}
\def\theequation{\arabic{section}.\arabic{equation}}

Having specified the calculational setup in
sec.~\ref{Sec:performance}, we are now in the position to present
results for the GT matrix element using the SMS 2N interactions of
Ref.~\cite{Reinert:2017usi}, along with the leading
contributions to the 3NF and two-body current operators.  

\subsection{The GT matrix element using the SMS chiral 2N forces and 1N currents}
\label{2NFonly}

We start with the single-nucleon 
contributions to the GT
matrix elements calculated using the two-nucleon forces (2NFs) from
Ref.~\cite{Reinert:2017usi} without 3NF. 
In Table \ref{tab:2NF}, we show the corresponding results 
for all orders and cutoff
values of $\Lambda = 400$, $450$, $500$ and $550$~MeV. 
\begin{table*}[t]
  \begin{ruledtabular}
    \begin{tabular*}{\textwidth}{@{\extracolsep{\fill}}lrrrr}
 & $\Lambda = 400$~MeV     & $\Lambda = 450$~MeV    &
                                                                  $\Lambda = 500$~MeV    & $\Lambda = 550$~MeV    \\ \hline  
      2NF at LO & $96.73$ & $96.15$ & $95.45$ & $94.64$ \\
      2NF at NLO & $94.52$ & $93.99$ & $93.52$ & $93.04$ \\
      2NF at N$^2$LO & $93.88$ & $93.08$ & $92.28$ & $91.44$ \\
      2NF at N$^3$LO & $93.63$ & $93.12$ & $92.64$ & $92.23$ \\
      2NF at N$^4$LO & $93.83$ & $93.32$ & $92.83$ & $92.44$ \\
      2NF at N$^4$LO$^+$ & $93.78$ & $93.23$ & $92.73$ & $92.31$           
   \end{tabular*}
\caption{\label{tab:2NF} The GT matrix element $\times 10^2$
  calculated using the single-nucleon current operator at N$^2$LO
  specified in sec.~\ref{sec:1Ncurrent} and the two-nucleon
  interactions from Ref.~\cite{Reinert:2017usi} for different chiral
  orders and cutoff values. For the axial current, only the first term
  on the right-hand side of Eq.~(\ref{AxialCurrent1NFinal}) is taken
  into account  as appropriate at N$^2$LO. 
}
\end{ruledtabular}   
\end{table*}
The results obtained using the high-precision interactions
at N$^4$LO$^+$ agree well with the corresponding values from 
other 2N interaction models. For example, the AV18 potential \cite{Wiringa:1994wb}
leads to ${\rm GT}  = 92.24 \times 10^{-2}$, while the
N$^3$LO potentials of Ref.~\cite{Entem:2003ft} yield ${\rm GT}  =
93.63 \times 10^{-2}$ and ${\rm GT}  = 93.22 \times
10^{-2}$ for $\Lambda = 500$~MeV and $\Lambda = 600$~MeV, respectively \cite{Baroni:2016xll}.

\subsection{Determination of the 3NF parameters at N$^2$LO}
\label{sec:3NF}

Starting from N$^2$LO, the results listed in Table \ref{tab:2NF}
become incomplete since one needs to take into account the 3NF and
exchange current operators, whose consistently regularized expressions
are currently only available at
N$^2$LO. To simplify the interpretation of the obtained results, we
follow the strategy of Ref.~\cite{LENPIC:2022cyu} and
employ the NN interactions at the highest available order N$^4$LO$^+$
in combination with the dominant (i.e., N$^2$LO) contributions to the
3NF and current operators. This allows us to exclude possible
distortions of the results caused by the inaccurate description of two-nucleon
scattering data at N$^2$LO. Cleary, the predicted values for the GT matrix
element are still expected to be valid only at the N$^2$LO accuracy level. 

The short-range part of the leading two-body axial current depends on
the LEC $D$, see Eq.~(\ref{RegularizedFinal}), which also contributes
to the 3NF. As explained in the introduction, we aim here at making
parameter-free predictions for tritium beta decay by fixing this
LEC in nucleon-deuteron scattering. The expression of the regularized
N$^2$LO 3NF used in this study is given in Eq.~(\ref{3NFN2LO}), see
also Ref.~\cite{Maris:2020qne}. For the LECs $c_i$ that enter both the expressions for the 3NF and exchange axial
current, we employ the central values from the Roy-Steiner equation
analysis of Ref.~\cite{Hoferichter:2015tha}  at NNNLO$^{NN}$ \footnote{These are the values used in the
  NN potentials of Ref.~\cite{Reinert:2017usi}  at N$^4$LO$^+$.}, $c_1 = -1.10$~GeV$^{-1}$,
$c_3 = -5.54$~GeV$^{-1}$
and $c_3 = 4.17$~GeV$^{-1}$,
corrected  to account for the finite shifts due to pion loop diagrams at N$^3$LO
in both the 3NF and exchange axial current \cite{Bernard:2007sp,Krebs:2016rqz}  $c_1 \to c_1 - g_A^2
M_\pi/(64 \pi F_\pi^2) \simeq c_1 - 0.13$~GeV$^{-1}$, $c_3 \to c_3 + g_A^4
M_\pi/(16 \pi F_\pi^2) \simeq c_3 + 0.89$~GeV$^{-1}$ and $c_4 \to c_4 - g_A^4
M_\pi/(16 \pi F_\pi^2) \simeq c_4 - 0.89$~GeV$^{-1}$:
\beq
\label{ci_original}
c_1 = -1.23 \mbox{ GeV}^{-1}, \quad \quad
c_3 = -4.65 \mbox{ GeV}^{-1}, \quad \quad
c_4 = 3.28 \mbox{ GeV}^{-1}.
\eeq

The only remaining parameters in the 3NF are the LECs $D$ and $E$,
which are usually expressed in terms of the corresponding 
dimensionless constants $c_D$ and $c_E$ via
\beq
\label{DefcDcE}
D = \frac{c_D}{F_\pi^2 \Lambda_\chi}, \quad \quad
E = \frac{c_E}{F_\pi^4 \Lambda_\chi},
\eeq
with $\Lambda_\chi = 700$~MeV. These LECs can be constrained from a
broad range of observables including ground state energies and radii
of light and medium-mass nuclei, three-nucleon scattering,
nucleon-$^4$He scattering as well as the tritium beta decay, see
Refs.~\cite{Epelbaum:2002vt,Nogga:2005hp,Gazit:2008ma,Piarulli:2017dwd,Lynn:2017fxg,Ekstrom:2015rta,Wesolowski:2021cni}
and references therein. Throughout this work, we follow the general strategy of the
LENPIC collaboration by striving to constrain the short-range part of
the nuclear Hamiltonian from experimental data on the lightest
possible systems
\cite{LENPIC:2018ewt,Maris:2020qne,LENPIC:2022cyu,Epelbaum:2019kcf,Epelbaum:2022cyo}. It
is customary to require a correct 
reproduction of the $^3$H binding energy (BE) by fixing  the
value of the LEC $c_E$ as a function of $c_D$
\cite{Epelbaum:2002vt,Nogga:2005hp,Gazit:2008ma,Piarulli:2017dwd,LENPIC:2018ewt,Maris:2020qne,LENPIC:2022cyu}. To
determine the remaining LEC $c_D$, a number of nucleon-deuteron
(Nd) scattering observables were considered in Ref.~\cite{LENPIC:2018ewt}. It was
found that the strongest constraint on $c_D$ is imposed by the 
high-precision RIKEN data on the unpolarized cross section for elastic
Nd scattering at $E_N = 70$~MeV \cite{Sekiguchi:2002sf} in the angular range corresponding to
the cross-section minimum. 
\begin{figure}[t]
    \centering
    \includegraphics[width=0.75\linewidth]{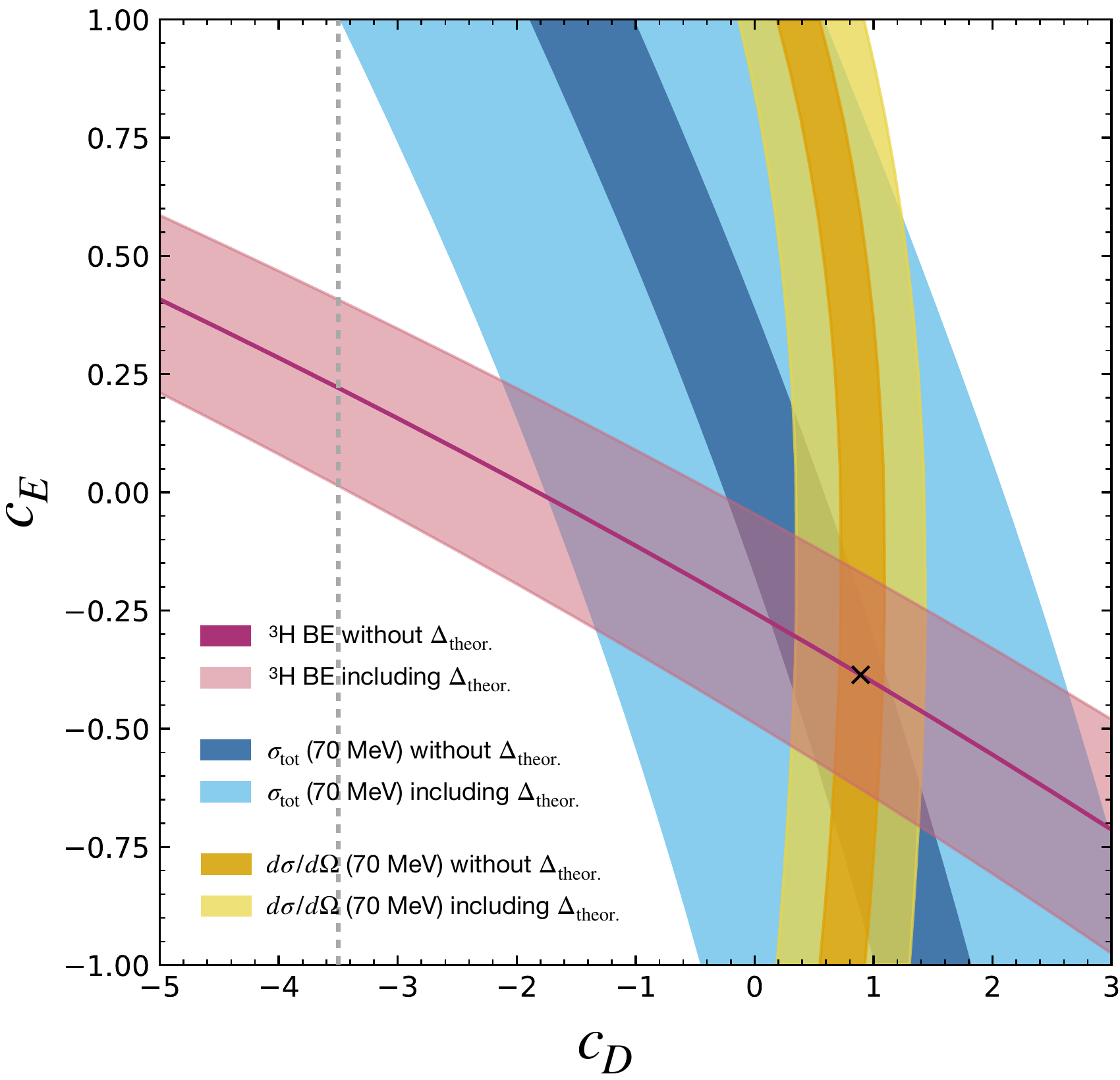}
    \caption{(Color online). Constraints on the values of $c_D$ and
      $c_E$ from fits to the $^3$H binding energy (red bands), Nd
      differential cross section data of Ref.~\cite{Sekiguchi:2002sf} at $E_N = 70$~MeV in the angular
      range of $\theta_{\rm cms} = 107^\circ\text{-}141^\circ$
      (yellow bands) as well as the Nd total cross
      section at $E_N = 70$~MeV from
      Ref.~\cite{Abfalterer:1998zz} (blue bands). Dark-shaded bands show the
      results of the fits using only the experimental uncertainties,
      while light-shaded bands also include N$^2$LO truncation errors
      estimated using the Bayesian model $\bar C_{0.5-10}^{650}$
      specified in  Ref.~\cite{Epelbaum:2019zqc} (the two types of errors
      are added in quadrature). All calculations are performed using the
      SMS N$^4$LO$^+$ NN force from Ref.~\cite{Reinert:2017usi} and the cutoff value
      of $\Lambda = 450$~MeV. The cross marks the values of $c_D$ and
      $c_E$ employed by the LENPIC collaboration in Ref.~\cite{LENPIC:2022cyu}. The gray dashed line
      shows the approximate constraint from the $^3$H $\beta$ decay as
      explained in sec.~\ref{sec:predictions}.
    }
    \label{fig:fig1}
  \end{figure}
  
In Fig.~\ref{fig:fig1}, we show the constraints imposed on $c_D$ and
$c_E$ from single-observable fits to the $^3$H BE as well
as the Nd differential and total cross section at $E_{N} =
70$~MeV. Our results are consistent with the well-known sensitivity of
the Nd differential cross-section minimum at
intermediate energies to the strength of the 3NF \cite{Gloeckle:1995jg,Witala:1998ey}.
Remarkably, the value of the LEC $c_E$ appears to be largely
insensitive to the differential cross section. Accordingly, the $^3$H BE and the Nd 
cross-section minimum at $70$~MeV are found to provide largely independent constraints on
$c_D$ and $c_E$, and these constraints are also consistent with those
imposed by the total cross section at the same energy. These
findings complement the results of Ref.~\cite{Wesolowski:2021cni},
where the
sensitivity of $c_D$, $c_E$ to the $^3$H and $^4$He BE and radii,
as well as the $^3$H $\beta$ decay was explored. Out of these observables,
only the $^3$H
$\beta$ decay was found to provide non-degenerate constraints on $c_D$
and $c_E$.
Our results show
that the Nd cross section minimum is well suited for 3NF parameter
fixing as an alternative to the $^3$H $\beta$ decay and 
provide strong support for the LENPIC fitting protocol of $c_D$, $c_E$
from the  $^3$H BE and Nd differential cross-section minimum at
$70$~MeV \cite{LENPIC:2018ewt,Epelbaum:2019zqc,Maris:2020qne,LENPIC:2022cyu}.
The robustness of this fitting strategy is further supported by
the generally good description (within errors) of experimental data for a broad range of Nd scattering
observables and properties of light p-shell nuclei found in
Refs.~\cite{Epelbaum:2019zqc,LENPIC:2022cyu,Skibinski:2023nnn,Endo:2024cbz}. 

In the first row of Table  \ref{tab:cDcE}, we list the central values of
the LECs $c_D$, $c_E$ determined from the $^3$H BE and the differential
cross-section minimum in elastic Nd scattering at
$70$~MeV for all considered cutoff values. We emphasize that 
few-nucleon LECs including $c_D$ and $c_E$ are to be considered as bare parameters in
the employed chiral EFT framework, and their values strongly depend on
the choice of NN interaction, expansion order, regularization
procedure, employed subtractions, 
and cutoff choices. It is, therefore, generally not possible to
directly compare the numerical values of $c_D$, $c_E$ across different
calculations.

\begin{table*}[t]
  \begin{ruledtabular}
    \begin{tabular*}{\textwidth}{@{\extracolsep{\fill}}lrrrr}
 & $\Lambda = 400$~MeV     & $\Lambda = 450$~MeV    & $\Lambda = 500$~MeV    & $\Lambda = 550$~MeV    \\ \hline  
 $c_D$ ($c_E$)   & $3.328$ ($-0.454$) & $0.892$ ($-0.386$) & $-1.279(-0.382)$  & $-3.626$ ($-0.410$)  \\
 $c_D$ ($c_E$)  using unsubtracted 3NF ($C=0$) & $5.208 \phantom{-}$  ($0.723$) & $2.756 \phantom{-}$ ($0.369$) & $0.520$ ($-0.014$) & $-2.025$ ($-0.503$)\\
  $c_D$ ($c_E$)  using effective values $c_i^{\rm eff}$ in the 3NF &     $5.479$ ($-0.538$) & $3.643$ ($-0.498$) & $2.346$ ($-0.547$) & $1.208$ ($-0.670$)
   \end{tabular*}
   \caption{\label{tab:cDcE} Values of the LECs $c_D$ and $c_E$
     determined from the $^3$H binding energy and Nd elastic
     scattering at $E_N = 70$~MeV as explained in the text. The
     results using our standard convention for the 3NF with additional
   subtractions are shown in the first row, while those using the
   unsubtracted 3NF expressions with $C=0$ are given in the second row. The
   results in the last row correspond to the reduced (effective)
   values of the LECs $c_{1,3,4}$ as explained in sec.~\ref{sec:effective}.}
\end{ruledtabular}   
\end{table*}

\subsection{N$^2$LO predictions for the GT matrix element}
\label{sec:predictions}

With the values of the LEC $c_D$ being determined as described in the
previous section, the expressions for the axial current at N$^2$LO are fixed in a
parameter-free way. 
In Table \ref{tab:full}, we show the resulting N$^2$LO predictions for
the GT matrix element for different cutoff values.  
\begin{table*}[t]
  \begin{ruledtabular}
    \begin{tabular*}{\textwidth}{@{\extracolsep{\fill}}lrrrr}
 & $\Lambda = 400$~MeV     & $\Lambda = 450$~MeV    &
                                                                  $\Lambda = 500$~MeV    & $\Lambda = 550$~MeV    \\ \hline  
      2NF at N$^4$LO$^+$ & $93.78$ & $93.23$ & $92.73$ & $92.31$
      \\
     2NF at N$^4$LO$^+$ $+$ 3NF & $93.87$ & $93.33$ & $92.86$ & $92.50$           \\
     2NF at N$^4$LO$^+$ $+$ 3NF  $+$ MEC & $103.41$ & $101.19$ & $99.63$ & $98.37$      
   \end{tabular*}
\caption{\label{tab:full} N$^2$LO predictions for the GT matrix
  element $\times 10^2$ for different cutoff values. The first row shows the incomplete results 
  calculated using the N$^4$LO$^+$ 2NF and the
  single-nucleon current, see sec.~\ref{2NFonly} for details. The
  values in the second row are obtained by taking into account the
  3NF contributions to the $^3$H and $^3$He wave functions. In the
  last row, the complete N$^2$LO results are given, which also incude the
  contribution from the two-body current in
  Eq.~(\ref{RegularizedFinal}). The employed values of the LECs $c_D$, $c_E$
  are given in the first row of Table \ref{tab:cDcE}. 
}
\end{ruledtabular}   
\end{table*}
As already mentioned above and also visible from Table \ref{tab:2NF}, the GT ME is fairly
insensitive to details of the interactions used to generate the wave
functions. In line with this observation, adding the N$^2$LO 3NF is
found to change the GT ME only by $\sim 0.1\, \text{-}\, 0.2\%$, depending on the
cutoff value.  Our results in the second row of Table \ref{tab:full}
agree well with those based
on high-precision phenomenological models\footnote{The one-body
  contributions to the GT ME using 11 phenomenological potential
  models with and without 3NFs were
  found in Ref.~\cite{Schiavilla:1998je}
  to lie in the range of ${\rm GT} = (92.2 \ldots
  93.7) \times 10^{-2}$.}. In contrast, the two-body 
contributions of $\sim 6\, \text{-}\, 10\%$ turn out to be surprisingly large
compared to the amount of underprediction of ${\rm GT}_{\rm emp} = 94.84 (19)
\times 10^{-2}$ based on the single-nucleon current. They lead
to a considerable overestimation of the GT ME at N$^2$LO, which becomes
particularly striking for the softest cutoff choice of $\Lambda = 400$~MeV. 
To judge upon the significance of the observed discrepancy, it is
important to estimate the truncation uncertainty of the N$^2$LO
predictions. Using the Bayesian model $\bar C_{0.5-10}^{650}$ of
Ref.~\cite{Epelbaum:2019zqc}, see also
Refs.~\cite{Furnstahl:2015rha,Melendez:2017phj} for pioneering applications of Bayesian methods to
uncertainty quantification in nuclear chiral EFT, we estimate the
N$^2$LO truncation errors from the order-by-order convergence
pattern for the GT ME $\times 10^2$ to be $2.3$, $1.9$, $1.7$ and
$1.5$ for $\Lambda = 400$, $450$, $500$ and $550$~MeV, respectively. 
However, the $^3$H $\beta$ decay  exhibits an irregular convergence
pattern with a nearly vanishing NLO correction, which puts the above
estimations into question, see Ref.~\cite{Wesolowski:2021cni} for
a related discussion and a similar conclusion.
This issue becomes further exacerbated by
the fact that the theoretical predictions for this quantity are
available only up through N$^2$LO. As will be argued below, there are
strong indications that the actual N$^2$LO truncation uncertainties
are considerably larger than the values quoted above. 

To gain more insights into the surprisingly large effect of the
two-body current, $\delta  {\rm GT}_{\rm MEC}$, it is instructive to look at its individual
contributions stemming from the long-range terms proportional to the
LECs $c_3$, $c_4$ and the short-range piece driven by
$c_D$:
\beq
\label{linearized}
\delta  {\rm GT}_{\rm MEC}  \; \approx \; c_3 \, \delta
{\rm GT}_{c_3} +
c_4 \, \delta
{\rm GT}_{c_4} +
c_D \, \delta
{\rm GT}_{c_D} \,,
\eeq
where $\delta {\rm GT}_{c_i}$ are assumed to be
independent of $c_{3,4,D}$. This approximate relationship neglects
nonlinear effects due to the dependence of the $^3$H and
$^3$He wave functions on $c_{3,4,D}$, but we will show that it is
valid to a very good accuracy. The individual contributions $\delta
{\rm GT}_{c_i}$ are given in Table \ref{tab:deltac}
for all cutoff values and both subtraction conventions.  
\begin{table*}[t]
  \begin{ruledtabular}
    \begin{tabular*}{\textwidth}{@{\extracolsep{\fill}}lcccc}
 & $\Lambda = 400\text{ MeV}$ & $\Lambda = 450\text{ MeV}$ & $\Lambda = 500\text{ MeV}$ & $\Lambda = 550\text{ MeV}$ 
 \\\hline  
$10^2 \times \delta
{\rm GT}_{c_3}\ \big(10^2 \times \delta
{\rm GT}_{c_3}^{C=0}\big)\, $ [GeV] &   
 $-0.17\ (-1.53)$ & $-0.34\ (-1.75)$ & $-0.53\ (-1.93)$ & $-0.73\ (-2.06)$ \\
$10^2 \times \delta
{\rm GT}_{c_4}\ \big(10^2 \times \delta
{\rm GT}_{c_4}^{C=0}\big)\, $ [GeV]&    
 $\phantom{-}1.20\ (-1.50)$ & $\phantom{-}1.51\ (-1.29)$ &  $\phantom{-}1.82\ (-0.94)$ & $\phantom{-}2.12\ (-0.51)$ \\
$10^2 \times \delta
{\rm GT}_{c_D}$ & 
 $1.43$ & $1.43$ & $1.36$ & $1.25$ \\
\end{tabular*}
   \caption{\label{tab:deltac} Individual contributions of the $c_3$-,
     $c_4$- and $c_D$-terms to $\delta {\rm GT}_{\rm MEC}$ as defined in Eq.~(\ref{linearized}). The values  in the
     brackets in the
     first and second rows correspond to the convention without
     subtractions, i.e., with $C=0$. The values are obtained using
     the $^3$H, $^3$He wave functions based on the N$^4$LO$^+$ NN
     potentials and the N$^2$LO 3NFs described in sec.~\ref{sec:3NF}.}
\end{ruledtabular}   
\end{table*}
Using the numerical values of $c_{3,4}$ quoted in
Eq.~(\ref{ci_original}) and of $c_D$ given in the first row of Table
\ref{tab:cDcE}, along with the individual
contributions $\delta {\rm GT}_{c_i}$ from Table
\ref{tab:deltac}, we obtain the decomposition of $\delta
{\rm GT}_{\rm MEC} $  as follows:
\beqa
\label{MECsubtracted}
\Lambda &=& 400\mbox{ MeV}: \quad \quad 10^2 \times \delta
{\rm GT}_{\rm MEC} 
\; =
\; 0.79_{c_3} \; +\; 3.94_{c_4} \; + \; 4.76_{c_D} \; = \;
9.49 \,, \nn
\Lambda &=& 450\mbox{ MeV}: \quad \quad 10^2 \times \delta
{\rm GT}_{\rm MEC}  
\; =
\; 1.58_{c_3} \; +\; 4.95_{c_4} \; + \; 1.28_{c_D} \; = \;
 7.81\,, \nn
\Lambda &=& 500\mbox{ MeV}: \quad \quad 10^2 \times \delta
{\rm GT}_{\rm MEC} 
\; =
\; 2.46_{c_3} \; +\; 5.97_{c_4} \; - \; 1.74_{c_D} \; = \;
 6.69\,, \nn
\Lambda &=& 550\mbox{ MeV}: \quad \quad 10^2 \times \delta
{\rm GT}_{\rm MEC} 
\; =
\; 3.39_{c_3} \; +\; 6.95_{c_4} \; - \; 4.53_{c_D} \; = \;
 5.81 \,. 
 \eeqa
By comparing these results with the difference between the values
quoted in the third and second rows of Table \ref{tab:full}, we
conclude that the approximate relation in Eq.~(\ref{linearized}) holds
at the $\sim 1\%$ accuracy level for all considered cutoff choices. 
Using our standard convention with the subtraction terms $\propto C$,  the
numerically largest contributions to $\delta
{\rm GT}_{\rm MEC} $ are generated by the $c_4$ and
$c_3$-operators. While  the long-range $c_3$ and $c_4$
terms add up coherently for all considered cutoff values, the $c_D$-contribution
is strongly $\Lambda$-dependent (as a consequence of the $\Lambda$-dependence
of the LEC $c_D$, see Table \ref{tab:cDcE}), and leads to sizable
cancellations for the harder cutoff
choices. Importantly, we observe that the individual contributions to $\delta
{\rm GT}_{\rm MEC} $ at N$^2$LO appear to be much larger in magnitude than
the total MEC contribution needed to reproduce the empirical value of  ${\rm GT}_{\rm emp} = 94.84 (19)
\times 10^{-2}$. This suggests a considerable degree of fine tuning
and points towards a sizable truncation
uncertainty for the GT ME at N$^2$LO. 

We can also use the approximate relationship in Eq.~(\ref{linearized}) to extract
the LEC $c_D$ needed to describe the empirical value of
the GT ME. Neglecting the dependence of the ME of the single-nucleon current on the
LEC $c_D$ and using the values in Tables \ref{tab:full} and
\ref{tab:deltac}, we infer for the cutoff $\Lambda = 450$~MeV the
constraint
$c_D \approx -3.5$, which is plotted as a gray vertical line in
Fig.~\ref{fig:fig1}.\footnote{Notice that in
  Ref.~\cite{Wesolowski:2021cni}, the constraint from the $^3$H
  $\beta$ decay is found to
  be nearly independent of $c_E$, which further supports
  the validity of our approximate considerations.}

\subsection{Consistency checks}
\label{sec:consistency}

Given the unexpected results reported in the previous section, it is important
to assess the robustness of the obtained predictions. Below, we perform
several consistency checks and quantify the impact of selected higher-order
contributions. 

\subsubsection{Comparison with the results by Baroni {\it et al.}}

To reduce the possibility of errors in the treatment of the
MEC, the most computationally involved part of our calculations, it is
instructive to compare the individual contributions $\propto
c_{3,4,D}$ with the values quoted in the literature. We stress that given the
different regularization schemes for the current operator and
different interaction models used to generate the tri-nucleon wave
functions, such a comparison is only expected to be meaningful at a
qualitative  level. 

We first use the results obtained by Baroni {\it et al.} in
Ref.~\cite{Baroni:2016xll}, who  show in their Table I the contributions of the
long-range MEC to the GT ME using the values of $c_3 = -3.20$~GeV$^{-1}$,
$c_4 = 5.40$~GeV$^{-1}$ and $c_3 = -5.61$~GeV$^{-1}$,
$c_4 = 4.26$~GeV$^{-1}$, which are labeled as N3LO(OPE)
and N3LO$^\star$(OPE), respectively. The corresponding contributions
to the GT ME are given for two Hamiltonian models: The one consisting
of the N$^3$LO NN potentials from Ref.~\cite{Entem:2003ft} along with the N$^2$LO 3NF,
and the one based on the combination of the AV18 NN potential \cite{Wiringa:1994wb}
with the UIX 3NF \cite{Pudliner:1995wk}. The two-body current operator is regularized
with a local cutoff in momentum space, without subtractions, using two
values of $\Lambda = 500$~MeV and $\Lambda = 600$~MeV (chosen to
coincide with the values used in the chiral EFT Hamiltonian). Notice
that the authors of Ref.~\cite{Baroni:2016xll} also include the leading $1/m$
contributions to the MEC, which are counted as higher-order
corrections in our scheme. Assuming that the
nonlocal relativistic correction in Eq.~(14) of Ref.~\cite{Baroni:2016xll},
suppressed by the factor of $(2m)^{-1}$, yields a small
contribution to the GT ME,  we can
extract the individual contributions $\delta
{\rm GT}_{c_3}$ and $\delta
{\rm GT}_{c_4}$ using the approximate relationship in
Eq.~(\ref{linearized}) from the values quoted in Table I of
Ref.~\cite{Baroni:2016xll} for the two sets of LECs $c_3$, $c_4$. By
solving the system of two linear equations, we obtain
\beqa
10^2 \times \delta {\rm GT}_{c_3}^{\rm N3LO/N2LO} \;=\;
-1.68 \, (-2.12) \text{ GeV} \quad &\text{for}& \Lambda \; = \; 500
\, (600)\text{ MeV}, \nn
10^2 \times \delta {\rm GT}_{c_4}^{\rm N3LO/N2LO} \;=\;
-0.80 \, (-1.19) \text{ GeV} \quad &\text{for}& \Lambda \; = \; 500
\, (600)\text{ MeV},
\eeqa
for the EFT-based Hamiltonian and 
\beqa
10^2 \times \delta {\rm GT}_{c_3}^{\rm AV18/UIX} \; =\;
-2.32 \, (-2.66) \text{ GeV} \quad &\text{for}& \Lambda \; = \; 500
\, (600)\text{ MeV},\nn
10^2 \times \delta {\rm GT}_{c_4}^{\rm AV18/UIX} \; =\;
-1.08 \, (-0.18) \text{ GeV} \quad &\text{for}& \Lambda \; = \; 500
\, (600)\text{ MeV},
\eeqa
for the phenomenological combination of the AV18 NN potential with the
UIX 3NF. These values agree reasonable well, given scheme dependence,
with our results quoted in Table \ref{tab:deltac}:
\beqa
10^2 \times \delta {\rm GT}_{c_3}^{C=0} \;=\;
-1.53 \ldots -2.06 \text{ GeV} \quad &\text{for}& \Lambda \; = \; 400
\ldots 550\text{ MeV}, \nn
10^2 \times \delta {\rm GT}_{c_4}^{C=0} \;=\;
-1.50 \ldots -0.51 \text{ GeV} \quad &\text{for}& \Lambda \; = \; 400
\ldots 550\text{ MeV}.
\eeqa

To benchmark the $c_D$-contribution, we use the results of
Ref.~\cite{Baroni:2018fdn}. This study uses four
different versions of the Norfolk two- and three-body interactions (Ia, Ib, IIa
and IIb) \cite{Piarulli:2016vel}, which differ from each other by the employed cutoff values and
fitting strategies of the LECs entering the NN potential. Using the
values of the LEC $c_D$ and the corresponding contributions to the GT
ME labeled N3LO(CT), which are given in Table I of Ref.~\cite{Baroni:2018fdn}, we extract $\delta
{\rm GT}_{c_D}^{\rm Norfolk}$ after conversion to our
convention\footnote{The quoted N3LO(CT) values take into account not
  only the $c_D$ contribution, but also the short-range parts of the
  one-pion exchange 2N axial current (which are similar to our
  subtraction terms), see Eq.~(3.2) of Ref.~\cite{Baroni:2018fdn}. In
  addition, that paper uses a different value of the scale
  $\Lambda_\chi$, namely $\Lambda_\chi = 1$~GeV, when relating the LECs $D$
  and $c_D$ according to Eq.~(\ref{DefcDcE}).}
\beqa
10^2 \times \delta {\rm GT}_{c_D}^{\rm Norfolk} \;=\;
1.11 \ldots 1.27 \quad &&\text{for the set of models Ia, Ib, IIa, IIb}.
\eeqa
Again, these values compare reasonably well with our result 
\beqa
10^2 \times \delta {\rm GT}_{c_D} \;=\;
1.25 \ldots 1.43 \quad &&\text{for} \; \; \Lambda = 400 \ldots 550\text{ MeV}.
\eeqa

Thus, while the individual contributions $\delta {\rm GT}_{c_{3,4,D}}$ to the GT ME clearly exhibit a significant
dependence on the employed regularization and wave functions, we
conclude that our numerical results for all considered terms are consistent with the
values reported in Refs.~\cite{Baroni:2016xll,Baroni:2018fdn}.

\subsubsection{Removing subtractions}

To verify the robustness of our predictions for the GT ME, we have redone the
calculations described in sec.~\ref{sec:predictions} using the expressions for the
2N axial current and 3NF without additional subtractions, i.e., by
setting $C=0$. As already emphasized, these subtractions merely
represent a convention to reshuffle short-range pion-exchange
interactions into contact terms allowed by power counting. Since the
long-range (short-range) contributions are regularized using the local
(nonlocal) Gaussian cutoffs, the subtracted and unsubtracted
interactions are only equivalent to each other up to N$^4$LO and
higher-order terms. We further emphasize that the $c_D$-contribution
gets subtracted in the 3NF, see Eq.~(\ref{3NFN2LO}), but not in the current operator (due to the
absence of the pion propagator). Comparing the predictions obtained using two different
conventions, therefore, provides a rather nontrivial consistency check of our
results, in particular since switching off the subtractions strongly affects
the values of the LECs $c_D$ and $c_E$ as shown in the first and second rows
of Table \ref{tab:cDcE}. Using Eq.~(\ref{linearized}) and plugging in the
corresponding values from Table \ref{tab:deltac}, along with the
$c_D$-values listed in the second row of Table \ref{tab:cDcE}, we read
out the MEC contributions to the GT ME:
\beqa
\label{MECWithoutSubtractions}
\Lambda &=& 400\mbox{ MeV}: \quad \quad 10^2 \times \delta
{\rm GT}_{\rm MEC} 
\; =
\; 7.11_{c_3} \; -\; 4.92_{c_4} \; + \; 7.45_{c_D} \; = \;
9.64 \,, \nn
\Lambda &=& 450\mbox{ MeV}: \quad \quad 10^2 \times \delta
{\rm GT}_{\rm MEC}  
\; =
\; 8.14_{c_3} \; -\; 4.23_{c_4} \; + \; 3.94_{c_D} \; = \;
 7.85\,, \nn
\Lambda &=& 500\mbox{ MeV}: \quad \quad 10^2 \times \delta
{\rm GT}_{\rm MEC} 
\; =
\; 8.97_{c_3} \; -\; 3.08_{c_4} \; + \; 0.71_{c_D} \; = \;
 6.60\,, \nn
\Lambda &=& 550\mbox{ MeV}: \quad \quad 10^2 \times \delta
{\rm GT}_{\rm MEC} 
\; =
\; 9.58_{c_3} \; -\; 1.67_{c_4} \; - \; 2.53_{c_D} \; = \;
 5.38 \,. 
 \eeqa
By comparing these values with those given in Eq.~(\ref{MECsubtracted}), we conclude that
both subtracted and unsubtracted versions of the 3NF and 2N current
operator yield very similar results, as expected by consistency
arguments.

\subsubsection{Effective $c_i$'s}
\label{sec:effective}

Since the long-range components of the MEC generate the bulk of $\delta
{\rm GT}_{\rm MEC}$ (especially for hard cutoff choices), one may 
speculate that the observed overestimation of the GT ME might be related to
the inappropriate choice of numerical values of the subleading
$\pi$N LECs $c_{3,4}$. Indeed, from matching chiral perturbation
theory (ChPT) to the Roy-Steiner-equation solution for the $\pi$N amplitude at the
subthreshold point \cite{Hoferichter:2015tha,Siemens:2016jwj}, which
is believed to provide  the most
reliable determination of $\pi$N LECs, it is known that the numerical
values of $c_{3,4}$ exhibit a significant dependence on the
EFT expansion order. This is, of course, not surprising in view of the well-known
slow convergence of heavy-baryon ChPT in the single-nucleon sector,
see Ref.~\cite{Ekstrom:2025ost} for a related recent discussion, and has
implications for the long-range 3NF with the $\pi$N
scattering amplitude entering as a subprocess. In Ref.~\cite{Krebs:2012yv},
it was found by considering the pion-pole contributions to the
$2\pi$-exchange 3NF, that the leading (i.e., N$^3$LO) and subleading
(i.e., N$^4$LO) loop contributions can be well approximated by
changing the values of the LECs $c_{3,4}$ in the corresponding
tree-level expression at N$^2$LO. By matching the full N$^4$LO result
for the structure functions $A(q_2)$ and $B(q_2)$ entering  the
$2\pi$-exchange 3NF to the tree-level expression, the ``effective''
values of $c_{1,3,4}$ were determined,
\beq
\label{EffectiveLECs}
c_1^{\rm eff} = -0.37 \mbox{ GeV}^{-1}, \quad \quad
c_3^{\rm eff} = -2.71 \mbox{ GeV}^{-1}, \quad \quad
c_4^{\rm eff} = 1.41 \mbox{ GeV}^{-1},
\eeq
which implicitly account for higher-order corrections to the longest-range 3NF
topology. The resulting smaller-in-magnitude values for these LECs are
consistent with the observations made in Refs.~\cite{Krebs:2013kha,Epelbaum:2014sea} from looking at the
corresponding $r$-space 3N potentials, which show that the loop corrections to
the $2\pi$-exchange 3NF topology tend to significantly reduce the
dominant tree-level predictions. 

It is, therefore, interesting to redo the analysis of the tritium $\beta$-decay using the
reduced effective values of the $\pi$N LECs. To this aim, we have repeated the
determination of $c_D$, $c_E$ from the $^3$H BE and cross section
minimum at $70$~MeV using $c_{1,3,4}^{\rm eff}$ from
Eq.~(\ref{EffectiveLECs}), see the last row of Table \ref{tab:cDcE}.
Using the approximate relation in Eq.~(\ref{linearized}) and the values quoted
in Table \ref{tab:deltac}, we obtain the following decomposition of $\delta
{\rm GT}_{\rm MEC} $ into the individual contributions:
\beqa
\Lambda &=& 400\mbox{ MeV}: \quad \quad 10^2 \times \delta
{\rm GT}_{\rm MEC} 
\; =
\; 0.46_{c_3} \; +\; 1.69_{c_4} \; + \; 7.83_{c_D} \; = \;9.98
 \,, \nn
\Lambda &=& 450\mbox{ MeV}: \quad \quad 10^2 \times \delta
{\rm GT}_{\rm MEC} 
\; =
\; 0.92_{c_3} \; +\; 2.13_{c_4} \; + \; 5.21_{c_D} \; = \; 8.26
\,, \nn
\Lambda &=& 500\mbox{ MeV}: \quad \quad 10^2 \times \delta
{\rm GT}_{\rm MEC}  
\; =
\; 1.44_{c_3} \; +\; 2.57_{c_4} \; + \; 3.35_{c_D} \; = \; 7.36
\,, \nn
\Lambda &=& 550\mbox{ MeV}: \quad \quad 10^2 \times \delta
{\rm GT}_{\rm MEC} 
\; =
\; 1.98_{c_3} \; +\; 2.99_{c_4} \; + \; 1.73_{c_D} \; = \; 6.70
 \,. 
 \eeqa
Using the smaller-in-magnitude effective values of the $\pi$N LECs, the contribution of the
long-range part of the 3N current operator to $\delta
{\rm GT}_{\rm MEC}$ gets reduced compared to Eq.~(\ref{MECsubtracted}), but this effect is
overcompensated by the increased values of $c_D$, so that the 
resulting predictions for  $\delta {\rm GT}_{\rm MEC} $ become even
$\sim 5 \ldots 15\%$ larger than those given in sec.~\ref{sec:predictions}. 

\subsubsection{Relativistic corrections to the single-nucleon current}

We have also estimated the size of the leading
relativistic corrections. All results presented above are obtained
using the static approximation for the single-nucleon axial current by
neglecting the last term in Eq.~(\ref{AxialCurrent1NFinal}), and assuming that the $^3$He nucleus in the final
 state has vanishing momentum. For the cutoff $\Lambda = 450$~MeV, the leading relativistic
 correction $\sim p^2/m^2$ from the last term in
 Eq.~(\ref{AxialCurrent1NFinal}) yields
 the contribution to the GT ME of $\delta
{\rm GT}_{p^2/m^2}  = -0.5 \times 10^{-2}$, which is more than
an order of magnitude smaller than the observed discrepancy. This
value agrees qualitatively with the results of Ref.~\cite{Baroni:2018fdn} using the
Illinois models for the nuclear Hamiltonian, see the results labeled
N2LO(RC) in their Table I. We have also verified that the $p/m$-contribution to
the single-nucleon vector-isovector current \cite{Krebs:2019aka} has a
small effect on the tritium half-life. The
size of higher-order relativistic corrections in Eq.~(\ref{AxialCurrent1N}) appears to be very
small, $\delta
{\rm GT}_{\rm higher-order\; rel.}  = 0.02 \times 10^{-2}$, while the effect of
the nonvanishing momentum of $^3$He is completely negligible. We thus
conclude that relativistic effects cannot explain 
the observed discrepancy in the predicted values of the GT ME. 

\subsubsection{Sensitivity to the NN off-shell short-range
  interactions at N$^3$LO}
\label{sec:OffShell}
 
Finally, to probe the impact of (some of the) truncated short-range contributions at
N$^3$LO, we have redone the analysis using phase-equivalent but
off-shell different N$^4$LO$^+$ NN potentials introduced in
Ref.~\cite{Epelbaum:2025aan}. These interactions employ different choices for the
off-shell behavior of the NN short-range interactions. Specifically,
the NN contact interactions at N$^3$LO involve three operators acting
in the $^1$S$_0$,  $^3$S$_1$ and $^3$S$_1$-$^3$D$_1$ channels, which
can be eliminated by means of suitably chosen unitary
transformations \cite{Reinert:2017usi}. The corresponding LECs cannot be
determined from  two-nucleon scattering data alone and have been set to zero
in Ref.~\cite{Reinert:2017usi}. In Ref.~\cite{Epelbaum:2025aan}, a set of 27  N$^4$LO$^+$ NN
potentials was generated for the cutoff $\Lambda = 450$~MeV, where
these LECs were set to fixed values of 
natural size. Upon refitting the remaining contact interactions, the
resulting family of 
27 potentials has been shown to provide a nearly identical description of
NN scattering data below pion production threshold, while featuring different off-shell behavior at the
N$^3$LO accuracy level. In Ref.~\cite{Epelbaum:2025aan}, these potentials were used to
quantify the intrinsic scheme dependence of 3NFs in chiral EFT by
considering 3N scattering and bound-state observables. Here,
we follow the same idea and use these phase-equivalent but off-shell
different NN potentials to probe the
impact of selected short-range N$^3$LO contributions on tritium
$\beta$ decay. To this aim, we performed an independent
determination of the LECs $c_D$ and $c_E$ for each interaction model
by using the fitting protocol explained in sec.~\ref{sec:3NF} and
calculated the GT ME as described in sec.~\ref{sec:predictions}. The
resulting predictions for the $^3$H GT ME using the cutoff $\Lambda = 450$~MeV
are found to lie in the range of ${\rm GT}  = (99.80
\ldots 104.32) \times 10^{-2}$. The spread in the predictions 
reflects the  impact of the considered selected N$^3$LO short-range
contributions of natural size and
may serve as a lower bound estimate of the N$^2$LO truncation
uncertainty.

\section{Discussion and conclusions} \label{Sec:summary}

In this paper we have carried out a detailed analysis of tritium
$\beta$ decay at N$^2$LO of the chiral expansion focusing, in
particular, on the role played by the two-body current. The main
conclusions of our study can be summarized as follows. 
\begin{itemize}
\item
We have shown that the $^3$H binding energy and Nd differential
cross section around its minimum at intermediate scattering energies provide independent constraints on
the LECs $c_D$ and $c_E$ entering the 3NF at N$^2$LO. This is in
contrast to the low-energy few-nucleon observables including the binding
energies and charge radii of the $A=3,4$ nuclei and the Nd doublet
scattering length, which are known to be strongly correlated due to
a universal behavior of systems with large scattering lengths
\cite{Braaten:2004rn}. Therefore, these observables alone do not provide sufficient
information to reliably pin down the values of $c_D$ and $c_E$
\cite{Gazit:2008ma,Wesolowski:2021cni}. In 
contrast, using the $^3$H binding energy in combination with the
high-precision experimental data on the Nd differential
cross section at $E_{N} = 70$~MeV from Ref.~\cite{Sekiguchi:2002sf}, as done
in Refs.~\cite{LENPIC:2018ewt,Maris:2020qne,LENPIC:2022cyu}, is shown to impose stringent constraints on $c_D$ and $c_E$,
which are consistent with the constraint placed by the 
Nd total cross section at $E_{N} = 70$~MeV. Notice further that the resulting
LENPIC Hamiltonian was already successfully applied to a broad range of Nd scattering
observables and spectra of light p-shell nuclei  in Ref.~\cite{LENPIC:2022cyu}, leading to a
generally good description of experimental data (within errors).  
\item
Using the value of $c_D$  determined from Nd scattering, we made
parameter-free predictions for tritium $\beta$ decay at the N$^2$LO
accuracy level. In line with the existing calculations by other groups, see, e.g., Refs.~\cite{Baroni:2016xll,Baroni:2018fdn,King:2020wmp}, we found that the
results based on the single-nucleon
current lead to a slight underestimation of the empirical
value of ${\rm GT}_{\rm emp} = 94.84 (19)
\times 10^{-2}$. In particular, using the N$^4$LO$^+$ SMS NN
potentials from Ref.~\cite{Reinert:2017usi} with the cutoff values between $\Lambda =
400$~MeV and $550$~MeV, we found
 ${\rm GT}  = (92.31 \ldots 93.78 )
\times 10^{-2}$  (with the smallest value corresponding to the largest
value of the momentum-space cutoff $\Lambda$). The inclusion of the N$^2$LO 3NF in the calculation of
the $A=3$ wave functions was found to have a tiny effect by changing these values to
 ${\rm GT}  = (92.50 \ldots  93.87 )
 \times 10^{-2}$. On the other hand, the 2N current operator at N$^2$LO
 is found to yield surprisingly large contributions, $\delta {\rm GT}_{\rm MEC}  = (5.81 \ldots 9.49)
\times 10^{-2}$ (with the smallest value corresponding to the hardest
cutoff choice), which lead to an overprediction of the GT matrix
element by the amount of $\sim 4 \ldots 9\, \%$ depending on the cutoff. 
\item
To  reduce the possibility of errors, we have benchmarked our
results for the contributions to the GT matrix element generated by
the individual terms in the 2N current operator
against the values available in the literature. In spite of 
different nuclear wave functions and different functional
forms and numerical values of the regulators, we found our results to
be consistent with those quoted in Refs.~\cite{Baroni:2016xll,Baroni:2018fdn}. Furthermore, to
assess robustness of our conclusions, we have repeated the analysis by
using the version of the 3NF and 2N current operator without
additional subtractions, i.e., by setting $C=0$ instead of using Eq.~(\ref{BIGC}).  
We have also investigated whether  the observed
overprediction of the GT ME can be traced back to the large numerical values of the $\pi$N
LECs $c_i$ at N$^2$LO by performing the calculations
using the reduced values from Ref.~\cite{Krebs:2012yv},
which effectively mimic the N$^3$LO and N$^4$LO loop effects in the
longest-range 3NF. Despite the observed (and expected) large
differences between the individual
contributions, we found these alternative calculations to yield
similar total predictions for the GT matrix element. 
\end{itemize}  
Given all above, we conclude that the $^3$H GT matrix element is
substantially overpredicted at N$^2$LO if the value of the LEC $c_D$
is extracted from Nd scattering. This
observation is not entirely unexpected, see, e.g.,
Ref.~\cite{King:2020wmp} for a related discussion.
The origin of the problem seems to be related to a fine-tuned nature of this
observable. Indeed,  the expected size of
two-body contributions to the $^3$H $\beta$ decay can be inferred from the
parameter-free long-range currents $\propto c_{3,4}$. While the
individual $c_3$- and $c_4$-contributions are strongly scheme- and regulator-dependent,
our results in Eqs.~(\ref{MECsubtracted}) and
(\ref{MECWithoutSubtractions}) show that the long-range MEC-effects can be as large as
$\delta {\rm GT}_{\rm MEC, \; long-range}  \sim 10 \times 10^{-2}$. Thus, to
reproduce the empirical value of the GT ME at N$^2$LO, the short-range
part of the MEC $\propto c_D$ needs to be fine tuned to largely cancel
the positive contribution of the long-range 2N axial current. Our
results show 
that the required fine-tuning between the long- and short-range components of
the axial currents is not realized at the N$^2$LO accuracy level. 

It is difficult to quantitatively assess the significance of the
observed discrepancy between our N$^2$LO predictions for the GT ME and
its empirical value. The irregular convergence pattern of chiral EFT
for this observable \cite{Wesolowski:2021cni} and the lack of theoretical predictions beyond
N$^2$LO prevent one from obtaining reliable estimations of the truncation
uncertainty using the standard Bayesian methodology \cite{Furnstahl:2015rha,Melendez:2017phj}.  On the
qualitative level, using the long-range MEC contributions to the GT ME
as an estimate of the typical size of N$^2$LO corrections and
assuming the expansion parameter of $\sim 1/3$ \cite{LENPIC:2022cyu,Millican:2025sdp}, one
may expect the neglected N$^3$LO contributions to be of the order of
$\delta {\rm GT}_{\rm N^3LO}  \sim 3
\times 10^{-2}$. This estimate is in line with the variation of our
predictions for the GT ME induced by changing the off-shell
behavior of the NN potential at 
N$^3$LO within a natural range, as described in
sec.~\ref{sec:OffShell}.

Assuming the validity of the above uncertainty estimate, a
resolution of the observed overprediction of the GT ME would require
large negative two-body contributions at or even beyond
N$^3$LO, pointing towards a slow convergence of the chiral expansion
for the exchange axial current. One can think of two possible
mechanisms for such contributions to emerge. First, the determination
of the LEC $c_D$ from Nd scattering observables may lead to
significantly different values upon taking into account the
corrections to the 3NF beyond N$^2$LO. Secondly, loop corrections
to the axial NN current operator may be numerically enhanced due to
the appearance of large dimensionless coefficients, see Refs.~\cite{Epelbaum:2024gfg,Epelbaum:2026brf}
for a related discussion. Indeed, the loop contributions to  
the one- and two-pion exchange NN axial current contributions are enhanced by one
power of $\pi$ relative to the expected suppression factor of $\sim M_\pi^2/(4
\pi F_\pi)^2$ \cite{Krebs:2016rqz}. Interestingly, these parameter-free corrections to the
exchange axial current were found in Refs.~\cite{Baroni:2015uza,Baroni:2018fdn} to generate
numerically large and negative contributions to the GT ME. In
particular, using the expressions from Ref.~\cite{Baroni:2015uza} derived in the
framework of time-ordered perturbation theory, the resulting
corrections were found to be $\delta {\rm GT}_{\rm N3LO, \; long-range}  = (-4.30
\ldots -7.57)  \times 10^{-2}$ in Ref.~\cite{Baroni:2015uza} and $\delta {\rm GT}_{\rm N3LO, \; long-range}  = (-6.71
\ldots -7.32)  \times 10^{-2}$ in Ref.~\cite{Baroni:2018fdn}, where the spread
in the quoted values reflects the cutoff and interaction dependence. Using the expressions
obtained in Ref.~\cite{Krebs:2016rqz} with the method of unitary transformation,
the authors of Ref.~\cite{Baroni:2018fdn} found somewhat smaller in magnitude but
still numerically large contributions of $\delta {\rm GT}_{\rm N3LO, \; long-range}  = (-3.64
\ldots -5.43)  \times 10^{-2}$. We emphasize, however, that these
results should only be regarded as indicative in view of the still
missing expressions for the axial current obtained using a symmetry-preserving
cutoff regularization. 

Our results have important implications for {\it ab initio}
calculations of electroweak processes involving nuclei in chiral
EFT. The apparent slow convergence of the chiral expansion
for the exchange current operator puts into question the
frequently employed approach for fixing the LEC $c_D$ using the empirical
value of the triton GT matrix element. While such an approach can, 
to some extent, be expected to yield a phenomenologically adequate description of weak
processes by effectively absorbing the (presumably numerically large)
contributions to the axial current operator beyond N$^2$LO into a redefinition
of $c_D$, probing weak currents in different kinematical conditions
through, e.g., muon capture reactions may suffer from large
uncertainties. In addition, as shown in Fig.~\ref{fig:fig1}, such fitting
strategy may result in a description of three-nucleon scattering
observables that is incompatible with experimental data.   
In any case, our findings suggest that the truncation uncertainty of chiral EFT
predictions for weak processes at the N$^2$LO accuracy level might be
significantly underestimated in existing calculations.

To clarify the situation with tritium $\beta$ decay and gain more
quantitative insights into the convergence pattern of chiral EFT for
exchange axial currents, it is necessary to extend this
study to N$^3$LO. Using nucleon-deuteron scattering data to fix the
short-range part of the 3NF would then still allow one to make
parameter-free predictions for the GT ME. Work along these lines is in progress 
using the recently proposed gradient flow formulation of chiral EFT \cite{Krebs:2023ljo,Krebs:2023gge},
which allows one to derive regularized expressions for the 3NF and
exchange currents in harmony with the chiral and gauge symmetries of
the Standard Model. At the N$^4$LO accuracy level, however, the correct
reproduction of the tritium $\beta$ decay is probably achievable in
a trivial way. Indeed, the large number of the new $D$-like
operators with undetermined LECs at this expansion order \cite{Huesmann:2026khj} suggests that the constraint
placed by the empirical value of the GT ME will likely be
imposable without compromising the description of other
observables. Using tritium $\beta$-decay to constrain the corresponding
LECs at N$^4$LO could then be efficiently implemented by applying the emulation technique
proposed in Ref.~\cite{Heihoff:2026ycq}.

\acknowledgments
We are grateful to all members of the LENPIC collaboration for sharing
their insights into the considered topics. 
		The work was also supported by the National Science Centre, 
		Poland under Grant IMPRESS-U 2024/06/Y/ST2/00135; 
		in part by the Excellence Initiative – Research University Program 
		at the Jagiellonian University in Krak\'{o}w,
		by the European Research Council (ERC) under the
                European Union's Horizon 2020 research and innovation
                programme (grant agreement No.~885150), by the MKW NRW
		under the funding code NW21-024-A, 
by JST ERATO (Grant No. JPMJER2304), by JSPS KAKENHI (Grant
No. JP20H05636), and by BMBF through the ErUM-Data project DEMOS.
Some numerical calculations were performed on the supercomputers at the JSC, J\"{u}lich, Germany.

\appendix
\renewcommand{\theequation}{\thesection.\arabic{equation}}

\section{Regularized expressions for the 3NF at N$^2$LO}
\label{appen:3NF}

The regularized expressions for the 3NF at N$^2$LO employed in our
study have the form
\beqa
\label{3NFN2LO}
V_{\rm 3N, \, reg.} &=& \frac{g_A^2}{8 F_\pi^4}\;  e^{- \frac{{\bm q}_1^2 +
M_\pi^2}{\Lambda^2}}\, e^{- \frac{{\bm q}_3^2 + M_\pi^2}{\Lambda^2}}
\bigg\{
\frac{\fet \sigma_1 \cdot {\bm q}_1  \; \fet \sigma_3 \cdot {\bm q}_3}{({\bm q}_1^{2} + M_\pi^2) \, ({\bm q}_3^{2} + M_\pi^2)}
\Big[\fet \tau_1 \cdot \fet \tau_3  \big( 2 c_3 \, {\bm q}_1 \cdot {\bm q}_3 - 4 c_1 M_\pi^2 \big)  
+  c_4 \fet \tau_1 \times \fet \tau_3 \cdot \fet \tau_2 \, {\bm q}_1 \times {\bm q}_3 
  \cdot \fet \sigma_2  \Big] \nn
&+&
C\, \frac{\fet \sigma_1 \cdot {\bm q}_1}{{\bm q}_1^{\, 2} + M_\pi^2}
\Big( 2 c_3 \, \fet \tau_1 \cdot \fet \tau_3\, \fet \sigma_3 \cdot {\bm q}_1 + c_4 \fet \tau_1 \times \fet \tau_3 \cdot \fet \tau_2 \; {\bm q}_1 \times \fet \sigma_3 
  \cdot \fet \sigma_2  \Big)   
  \nn
&+&
C  \, \frac{\fet \sigma_3 \cdot {\bm q}_3}{{\bm q}_3^{2} + M_\pi^2}
\Big( 2 c_3 \, \fet \tau_1 \cdot \fet \tau_3\, \fet \sigma_1 \cdot {\bm q}_3 + c_4 \fet \tau_1 \times \fet \tau_3 \cdot \fet \tau_2\; \fet \sigma_1 \times {\bm q}_3 
  \cdot \fet \sigma_2  \Big)  
  \nn
&+&
C^2 \, \Big( 2 c_3 \, \fet \tau_1 \cdot \fet \tau_3 \, \fet \sigma_1 \cdot \fet \sigma_3 + c_4 \fet \tau_1 \times \fet \tau_3 \cdot \fet \tau_2 \; \fet \sigma_1 \times \fet \sigma_3 
  \cdot \fet \sigma_2 \Big)  \bigg\}\nn
&-& \frac{g_A \, D}{8 F_\pi^2}\; \fet \tau_1 \cdot \fet \tau_3 \;
 e^{- \frac{{\bm p}_{12}^2 +{\bm p}_{12}^{\prime 2}
}{\Lambda^2}}\,
  e^{- \frac{{\bm q}_3^2 +
M_\pi^2}{\Lambda^2}}\, \bigg(
\frac{\fet \sigma_3 \cdot {\bm q}_3 }{{\bm q}_3^{2} + M_\pi^2} \; 
\fet \sigma_1 \cdot {\bm q}_3  + 
C  \, \fet \sigma_1 \cdot \fet \sigma_3 \bigg) \;  +\;   
\frac{1}{2} E \; \fet \tau_1 \cdot \fet \tau_2 \, e^{- \frac{{\bm p}_{12}^2 +{\bm p}_{12}^{\prime 2}
  }{\Lambda^2}}\, e^{- \frac{3 {\bm k}_{3}^2 +3 {\bm k}_{3}^{\prime 2}
}{4\Lambda^2}} \nn[4pt]
& +&  
\mbox{5 permutations}\,, 
\eeqa
where ${\bm q}_{i} = {\bm p}_i  ' - {\bm p}_i$ denote the momentum transfer of  nucleon $i$ with 
${\bm p}_i '$ and ${\bm p}_i$ being the corresponding final and
initial momenta, respectively. We have also introduced the Jacobi
momenta
\beq
{\bm p}_{12} = \frac{1}{2} ({\bm p}_1 - {\bm p}_2), \qquad
{\bm k}_3 = \frac{2}{3} \bigg({\bm p}_3 - \frac{1}{2} ({\bm p}_1 + {\bm p}_2 )\bigg)
\eeq
in the initial state and
\beq
{\bm p}_{12}^{\prime} = \frac{1}{2} ({\bm p}_1^{\prime} - {\bm p}_2^{
  \prime}), \qquad
{\bm k}_3^{\, \prime} = \frac{2}{3} \bigg({\bm p}_3^{ \prime} -
\frac{1}{2} ({\bm p}_1^{ \prime} + {\bm p}_2^{ \prime} )\bigg)
\eeq
in the final state. The subtraction constant $C$ is specified in
Eq.~(\ref{BIGC}). Notice that the pion-pole contributions to the
exchange axial current operator in Eq.~(\ref{Regularized}) can be directly read
off from the corresponding expression for the 3NF in
Eq.~(\ref{3NFN2LO}) by removing a single
pion-nucleon vertex, as explained in detail in Ref.~\cite{Krebs:2016rqz}.

\end{document}